\documentclass[aps, twocolumn, superscriptaddress, showpacs]{revtex4-1}
\pdfoutput=1

\usepackage{graphicx}
\usepackage{amssymb,amsmath}
\usepackage{bm}

\begin{document}

\title{Optimizing stability of mutual synchronization between a pair of limit-cycle oscillators with weak cross coupling}

\author{Sho Shirasaka}
\affiliation{Research Center for Advanced Science and Technology, University of Tokyo, Tokyo 153-8904, Japan}

\author{Nobuhiro Watanabe}
\affiliation{Department of Systems and Control Engineering,
Tokyo Institute of Technology, Tokyo 152-8552, Japan}

\author{Yoji Kawamura}
\affiliation{Department of Mathematical Science and Advanced Technology,
Japan Agency for Marine-Earth Science and Technology, Yokohama 236-0001, Japan}

\author{Hiroya Nakao}
\email{nakao@mei.titech.ac.jp (corresponding author)}
\affiliation{Department of Systems and Control Engineering,
Tokyo Institute of Technology, Tokyo 152-8552, Japan}
\affiliation{Department of Mechanical Engineering, University of California, Santa Barbara, California}

\begin{abstract}
We consider optimization of linear stability of synchronized
states between a pair of weakly coupled limit-cycle oscillators with
cross coupling, where different components of state variables of the oscillators
are allowed to interact.
On the basis of the phase reduction theory, the coupling matrix
between different components of the oscillator states
that maximizes the linear stability of the synchronized state
under given constraints on overall coupling intensity
and on stationary phase difference is derived.
The improvement in the linear stability is illustrated by using several
types of limit-cycle oscillators as examples.

\end{abstract}


\maketitle

\section{Introduction}

Synchronization of nonlinear oscillators is widely observed and often plays important functional roles in a variety of real-world systems~\cite{ref:pikovsky01,ref:strogatz03,strogatz15,ref:winfree80,Kuramoto,ref:hoppensteadt97,ref:ermentrout10,Kiss,Kiss2}.
Exploration of efficient methods for realizing stable synchronization between  coupled oscillators or between oscillators and driving signals is both fundamentally and practically important.
Improvement in the efficiency of collective synchronization in networks of coupled oscillators has been
extensively studied in the literature~\cite{ref:tanaka08,ref:yanagita10,ref:yanagita12,ref:yanagita14,ref:skardal14,ref:skardal16,ref:nishikawa06a,ref:nishikawa06b,ref:nishikawa10,Ravoori}
for both Kuramoto-type phase models and chaotic oscillators,
where optimization of coupling networks connecting the oscillators has been the main target.

In the analysis of synchronization dynamics between weakly coupled nonlinear oscillators undergoing
limit-cycle oscillations, the phase reduction theory has played a dominant role~\cite{ref:winfree80,Kuramoto,ref:hoppensteadt97,ref:ermentrout10,ref:brown04,ref:nakao16,ref:ashwin16}.
It allows us to simplify the dynamics of a pair of limit-cycle oscillators with weak coupling
to a simple scalar equation for their phase difference.
The phase reduction theory, originally developed for finite-dimensional smooth limit-cycle
oscillators, has recently been generalized to non-conventional limit-cycling systems such
as collectively oscillating populations of coupled oscillators~\cite{Kawamura1}, systems with time delay~\cite{Kotani,Novicenko1,Novicenko2},
reaction-diffusion systems~\cite{Nakao}, oscillatory fluid convection~\cite{Kawamura2}, and hybrid dynamical systems~\cite{Shirasaka}.
Recently, methods for optimizing periodic external driving signals for efficient injection
locking and controlling of a single nonlinear oscillator (or a population of uncoupled oscillators) have also been proposed
on the basis of the phase reduction theory~\cite{ref:moehlis06,ref:harada10,ref:dasanayake11,ref:zlotnik12,ref:zlotnik13,ref:pikovsky15,ref:tanaka14a,ref:tanaka14b,ref:tanaka15,ref:zlotnik2016,ref:hasegawa14a,ref:hasegawa14b}.
In this study, we consider a pair of coupled limit-cycle oscillators and try to optimize the linear stability of the synchronized state using the phase reduction theory.

In the analysis of mutual synchronization of coupled oscillators, linear diffusive coupling
between the oscillators is a common setup. However, in most cases, only the same
vector component of the state variables can interact between the oscillators and different
vector components of the oscillator states are usually not allowed to interact.
In this study, we analyze a pair of oscillators with weak cross coupling,
where different vector components of the oscillator states are allowed to interact,
that is, differences in each vector component of the oscillator states can be
feed-backed to every other component with a linear gain specified by a coupling matrix,
and optimize the coupling matrix so that the linear stability of the mutually synchronized state is maximized.

We use the phase reduction theory to simplify the dynamics of a pair of weakly coupled limit-cycle
oscillators to a scalar equation for the phase difference, and use the method
of Lagrange multipliers to derive the optimal coupling matrix for the cases with and without
frequency mismatch between the oscillators. Using three examples of simple limit-cycle oscillators, we illustrate
that the linear stability of the synchronized state is actually improved and also that
the stationary phase difference can be controlled by appropriately choosing the coupling matrix.

This paper is organized as follows: in Sec. II, we introduce the coupled-oscillator
model and derive the equation for the phase difference by using the phase reduction
theory. In Sec. III, we formulate the optimization problem for improving
linear stability of the phase-locked states. In Sec. IV, the theoretical results
are illustrated by several examples of limit-cycle oscillators.
Sec. V gives summary and discussion.

\section{Model}

In this section, we introduce a pair of nearly identical limit-cycle oscillators
with weak cross coupling, reduce the dynamical equations to coupled phase equations
by using the phase reduction theory~\cite{ref:winfree80,Kuramoto,ref:hoppensteadt97,ref:brown04,ref:ermentrout10,ref:nakao16,ref:ashwin16},
and derive the equation for the phase difference.

\subsection{A pair of cross-coupled oscillators}

We consider a pair of weakly and symmetrically coupled, nearly identical limit-cycle oscillators described by
\begin{align}
 \dot{{\bm X}}_1(t) &= {\bm F}_1({\bm X}_1) + \epsilon K ({\bm X}_2 - {\bm X}_1), \cr
 \dot{{\bm X}}_2(t) &= {\bm F}_2({\bm X}_2) + \epsilon K ({\bm X}_1 - {\bm X}_2),
 \label{model}
\end{align}
where ${\bm X}_1$ and ${\bm X}_2$ are the $m$-dimensional state vectors of the oscillators $1$ and $2$, respectively, ${\bm F}_1$ and ${\bm F}_2$ are $m$-dimensional vector-valued functions representing the dynamics of the oscillators, $K$ is a $m \times m$ matrix of coupling intensities between the components of the state variables, and $\epsilon$ is a small positive parameter $(0 < \epsilon \ll 1)$ indicating that the interaction is sufficiently small.

Here, although the oscillators are ``diffusively'' coupled, we assume that the
matrix $K$ is generally not diagonal and can possess non-diagonal elements.
That is, differences in each vector component of the oscillator states are returned to other components as feedback signals with appropriate gains.
Therefore, different components of the state variables of the oscillators can mutually interact. 
This gives the possibility to improve the stability of the synchronized state
by adjusting the non-diagonal elements of the coupling matrix, exceeding the
stability that is achievable only with the diagonal coupling.
We assume linear diffusive coupling in the following, but the argument can be straightforwardly generalized to nonlinear coupling; see Sec. V.

We assume that the properties of the oscillators are nearly identical and their difference is $O(\epsilon)$.
That is, the functions ${\bm F}_{1,2}$ can be split into a common part ${\bm F}$ and deviations ${\bm f}_{1,2}$ as
\begin{align}
{\bm F}_{1,2}({\bm X}) = {\bm F}({\bm X}) + \epsilon {\bm f}_{1,2}({\bm X}),
\end{align}
where ${\bm F}$, ${\bm f}_1$, and ${\bm f}_2$ are assumed to be $O(1)$.
We also assume that the common part of the oscillator dynamics, $\dot{\bm X}(t) = {\bm F}({\bm X})$,
possesses a stable limit-cycle solution ${\bm X}_0(t) = {\bm X}_0(t+T)$
of period $T$ and frequency $\omega = 2\pi / T$, and that the dynamics of the oscillator is only slightly deformed
and persists even if small perturbations from the deviations ${\bm f}_{1,2}$ and mutual coupling are introduced.
These assumptions are necessary for the phase reduction that we rely on in the present study.

\subsection{Phase reduction}

Under the above assumptions, we can simplify the dynamics of the coupled oscillators
to coupled phase equations by applying the phase reduction theory~\cite{ref:winfree80,Kuramoto,ref:hoppensteadt97,ref:brown04,ref:ermentrout10,ref:nakao16,ref:ashwin16}.
That is, we introduce a phase $\theta$ ($0 \leq \theta < 2\pi$) of the oscillator state near the limit-cycle solution
${\bm X}_0(t)$ that increases with a constant frequency $\omega$ in the absence of
perturbations, and represent the oscillator state on the limit cycle as a function of the phase $\theta(t)$ as ${\bm X}_0(\theta(t))$.

In the present case, we introduce phase variables $\theta_{1,2}$ of the two oscillators, represent the oscillator states near the limit-cycle orbit as ${\bm X}_{1,2}(t) = {\bm X}_0(\theta_{1,2}(t)) + O(\epsilon)$ as functions of $\theta_{1,2}(t)$ at $t$, and approximately describe their dynamics by using only $\theta_{1,2}$.
By following the standard phase reduction and averaging procedures, we can 
derive a pair of coupled phase equations, which is correct up to $O(\epsilon)$,  as
\begin{align}
 \dot{\theta}_1(t) &= \omega_1 + \epsilon \Gamma(\theta_1 - \theta_2), \cr
 \dot{\theta}_2(t) &= \omega_2 + \epsilon \Gamma(\theta_2 - \theta_1).
 \label{phasemodel}
\end{align}
The frequencies $\omega_{1,2}$ of the oscillators are given by
\begin{align}
\omega_{1, 2} &= \omega + \epsilon \frac{1}{2\pi} \int_0^{2\pi} {\bm Z}(\psi) \cdot {\bm f}_{1,2}({\bm X}_0(\psi)) d\psi \cr
&= \omega + \epsilon \langle {\bm Z}(\psi) \cdot {\bm f}_{1,2}({\bm X}_0(\psi))  \rangle_{\psi}
\label{freq}
\end{align}
and the phase coupling function $\Gamma(\phi)$ is given by
\begin{align}
\Gamma(\phi) 
&= \frac{1}{2\pi} \int_0^{2\pi} {\bm Z}(\phi + \psi) \cdot  K \{ {\bm X}_0(\psi) - {\bm X}_0(\phi + \psi) \} d\psi
\cr
&= \langle {\bm Z}(\phi + \psi) \cdot K \{ {\bm X}_0(\psi) - {\bm X}_0(\phi + \psi) \}  \rangle_{\psi}.
\label{phscpl0}
\end{align}
Here, we introduced an abbreviation for the average over phase from $0$ to $2\pi$,
\begin{align}
\langle A(\psi) \rangle_{\psi} = \frac{1}{2\pi} \int_0^{2\pi} A(\psi) d\psi,
\end{align}
where $A(\psi)$ is a $2\pi$-periodic function of $\psi$.
In the following, without loss of generality, we assume that $\omega_1 \geq \omega_2$,
and denote the frequency difference between the oscillators as $\epsilon \Delta \omega = \omega_1 - \omega_2 \geq 0$, where $\Delta \omega$ is $O(1)$.

The function ${\bm Z}(\theta)$ in Eqs.~(\ref{freq}) and (\ref{phscpl0}) is a phase sensitivity function of the limit cycle ${\bm X}_0(\theta)$ of the common part, $\dot{\bm X}(t) = {\bm F}({\bm X})$. It is given by a $2\pi$-periodic solution to the adjoint equation $\partial {\bm Z}(\theta) / \partial \theta = - J(\theta)^{\rm T} {\bm Z}(\theta)$, where $J(\theta)$ is a Jacobi matrix of the vector field ${\bm F}({\bm X})$ at ${\bm  X} = {\bm X}_0(\theta)$ and ${}^{\rm T}$ denotes the matrix transpose, and is normalized as ${\bm Z}(\theta) \cdot {\bm F}({\bm X}_0(\theta)) = \omega$.
By using the adjoint method by Ermentrout~\cite{ref:ermentrout10,ref:brown04,ref:nakao16}, i.e., by backwardly evolving the adjoint equation with occasional renormalization, ${\bm Z}(\theta)$ can be calculated numerically.

For convenience, we rewrite the phase coupling function as
\begin{align}
 \Gamma(\phi) 
&=
\langle {\bm Z}(\phi + \psi) \cdot K \{ {\bm X}_0(\psi) - {\bm X}_0(\phi+\psi) \}  \rangle_{\psi}
\cr
&=
{\rm Tr\ } ( K W(\phi)^{\rm T} ),
\end{align}
where
\begin{align}
W(\phi) = \langle {\bm Z}(\phi+\psi) \otimes \{ {\bm X}_0(\psi) - {\bm X}_0(\phi+\psi) \} \rangle_{\psi}
\label{linearW}
\end{align}
is a correlation matrix between the vector components of the phase sensitivity function and the state difference between the oscillators.
Here, the symbol $\otimes$ represents a tensor product
and ${\rm Tr}$ denotes the trace of a matrix.
See Appendix A for the definition and related matrix formulas.
Because ${\bm X}_0(\theta)$ and ${\bm Z}(\theta)$ are $2\pi$-periodic functions, 
$\Gamma(\phi)$ and $W(\phi)$ are also $2\pi$-periodic.

\subsection{Stability of the synchronized state}

From Eq.~(\ref{phasemodel}), the phase difference $\phi = \theta_1 - \theta_2$ (restricted to $-\pi \leq \phi \leq \pi$ hereafter) between the two oscillators obeys
\begin{align}
 \dot{\phi} = \epsilon \{ \Delta \omega + \Gamma_a(\phi) \},
 \quad
 \Gamma_a(\phi) = \Gamma(\phi) - \Gamma(-\phi).
 \label{phasedif0}
\end{align}
Here, $\Gamma_a(\phi)$ is the antisymmetric part of the phase coupling function $\Gamma(\phi)$; it is also
$2\pi$-periodic and satisfies $\Gamma_a(0) = \Gamma_a(\pm \pi) = 0$.
Therefore, if $\Delta \omega$ satisfies $- {\rm max}_\phi \Gamma_a(\phi) < \Delta \omega < - {\rm min}_\phi \Gamma_a(\phi)$, Eq.~(\ref{phasedif0}) has at least one stable fixed point at the phase differences satisfying
$\Delta \omega + \Gamma_a(\phi) = 0$.
We denote one of such fixed points as $\phi^*$.

From Eq.~(\ref{phasedif0}), the linear stability of $\phi^*$
is given by $\epsilon \Gamma_a'(\phi^*)$, where $\Gamma_a'(\phi^*)$
is the slope of $\Gamma_a(\phi)$ at $\phi = \phi^*$.
Thus, if a fixed point $\phi = \phi^*$ satisfies
\begin{align}
\Delta \omega + \Gamma_a(\phi^*) = 0
\end{align}
and
\begin{align}
\left. \Gamma_a'(\phi^*) = \frac{d}{d \phi} \Gamma_a(\phi) \right|_{\phi=\phi^*} < 0,
\label{const11}
\end{align}
the phase difference $\phi$ can take a stationary phase difference, and the two oscillators can mutually synchronize, or phase-locked to each other, with a stable phase difference $\phi = \phi^*$  (within the phase-reduction approximation).

By defining a new matrix
\begin{align}
V(\phi) = W(\phi) - W(-\phi),
\end{align}
the antisymmetric part of the phase coupling function and its slope can be expressed as
\begin{align}
 \Gamma_a(\phi) = {\rm Tr\ } \{ K V(\phi)^{\rm T} \},
 \quad
 \frac{d}{d \phi} \Gamma_a(\phi) = {\rm Tr\ } \{ K V'(\phi)^{\rm T} \},
\label{Kphasecoupling}
\end{align}
where $V'(\phi)$ represents derivative of $V(\phi)$ with respect to $\phi$.
Using $2\pi$-periodicity of ${\bm Z}(\theta)$ and ${\bm X}_0(\theta)$, the matrices $V(\phi)$ and $V'(\phi)$
can be expressed as
\begin{align}
V(\phi)
= \langle \{ {\bm Z}(\phi+\psi) - {\bm Z}(-\phi+\psi) \} \otimes {\bm X}_0(\psi) \rangle_{\psi},
\end{align}
and
\begin{align}
V'(\phi) &= W'(\phi) + W'(-\phi) \cr
& = \langle \{ {\bm Z}'(\phi+\psi) + {\bm Z}'(-\phi+\psi) \} \otimes {\bm X}_0(\psi) \rangle_{\psi},
\quad
\end{align}
where the derivative of $W(\phi)$ with respect to $\phi$ is given by
\begin{align}
W'(\phi) &= \frac{d}{d\phi} W(\phi)
= \langle {\bm Z}'(\phi+\psi) \otimes {\bm X}_0(\psi) \rangle_{\psi}.
\end{align}
See Appendix B for the calculations. We use these expressions in the next section. 

\section{Optimizing the coupling matrix}

In this section, we derive the optimal coupling matrix $K_{\rm opt}$ for stable synchronization (phase locking) of the two oscillators.

\subsection{Optimality condition and constraint on the coupling matrix}

Our aim is to maximize the linear stability of the synchronized (phase-locked) state, characterized by $- \epsilon \Gamma_a'(\phi^*)$,
by adjusting the coupling matrix $K$.
Other types of optimality conditions for synchronization have also been considered in the literature for nonlinear oscillators driven by periodic signals, such as maximization of frequency difference between the oscillator and signal for fixed coupling intensity~\cite{ref:harada10,ref:tanaka14a} and minimization of the phase diffusion constant under the effect of noise~\cite{ref:pikovsky15}, in addition to the maximization of linear stability~\cite{ref:dasanayake11,ref:zlotnik12,ref:zlotnik13} that we generalize to coupled oscillators in the present study~\cite{footnote}.

We first consider the simple case where the two oscillators are identical and share the same frequency, and optimize the stability of the {\em in-phase} synchronized state with zero phase difference, $\phi^* = 0$.
We then consider the general case with a frequency mismatch $\Delta \omega \geq 0$
and optimize the stability of the synchronized state with a {\em given} stationary phase difference $\phi^*$, which is not necessarily $0$.
In both cases, as a constraint on the overall connection intensity between the oscillators,
we fix the Frobenius norm (see Appendix A) of the coupling matrix $K$ as
$\| K \|^2 = P$, where $P > 0$ is a given constant. In the latter case, the stationary phase difference
$\phi^*$ is also constrained.

\subsection{Optimization for identical oscillators without frequency mismatch}

We first consider the simple case where the oscillators are identical, ${\bm F}_1 = {\bm F}_2$, and their frequencies are equal to each other, $\omega_1 = \omega_2 = \omega$ and $\Delta \omega = 0$.
In this case, the in-phase and anti-phase synchronized states $\phi^*=0$ and $\phi^*=\pi$ are always stationary solutions to Eq.~(\ref{phasedif0}) because $\Gamma_a(0) = \Gamma_a(\pm \pi) = 0$.

We thus try to find the coupling matrix $K$ that gives the maximum of linear stability of $\phi^* = 0$,
\begin{align}
- \epsilon \Gamma'_a(0),
\end{align}
subject to the constraint on the Frobenius norm of $K$,
\begin{align}
\| K \|^2 = P \quad (P>0).
\label{const0}
\end{align}
Because $\epsilon > 0$, we divide this quantity by $\epsilon$ and simply try to maximize
\begin{align}
- \Gamma'_a(0) = - \left. \frac{d}{d \phi} \Gamma_a(\phi)  \right|_{\phi=0},
\end{align}
which we also call ``linear stability'' for simplicity in the following.
We introduce an action,
\begin{align}
S( K, \lambda )
&= - \left. \frac{d}{d \phi} \Gamma_a(\phi)  \right|_{\phi=0} + \lambda \Big( \| K \|^2 - P \Big)  \cr
&= - {\rm Tr\ } \Big( K V'(0)^{\rm T} \Big) + \lambda \Big( \| K \|^2 - P \Big),
\end{align}
where $\lambda$ is a Lagrange multiplier.
The first term of $S$ represents the stability of the fixed point and the second term represents the constraint.

By differentiating $S$ by $K$ and $\lambda$, we obtain 
\begin{align}
\frac{\partial}{\partial K} S( K, \lambda )
=& 
- V'(0)+2 \lambda K
= 0
\end{align}
and the constraint
$\| K \|^2 = P$. Therefore, the optimal $K$ should satisfy
\begin{align}
K = \frac{1}{2 \lambda} V'(0).
\end{align}
Plugging this $K$ into Eq.~(\ref{const0}) yields
\begin{align}
\lambda = \pm \frac{1}{2\sqrt{P}} \| V'(0) \|.
\label{lambdapossibility}
\end{align}
It turns out that the negative sign should be chosen (see Eq.~(\ref{GoptcaseA})), so that the optimal coupling matrix is given by
\begin{align}
K_{\rm opt} = - \sqrt{P} \frac{V'(0)}{ \| V'(0) \| }.
\label{KoptcaseA}
\end{align}

The antisymmetric part of the phase coupling function with this $K_{\rm opt}$ is given, from Eq.~(\ref{Kphasecoupling}), by
\begin{align}
\Gamma_a(\phi) = {\rm Tr\ }\{ K_{\rm opt} V(\phi)^{\rm T} \} = - \frac{\sqrt{P} }{ \| V'(0) \| } {\rm Tr\ } \{ V'(0) V(\phi)^{\rm T} \}
\end{align}
and the optimal linear stability of the in-phase fixed point $\phi=0$ is given by
\begin{align}
- \Gamma_a'(0)
&= - {\rm Tr\ } \{ K_{\rm opt} V'(0)^{\rm T} \}
= \sqrt{P} \| V'(0) \|.
\label{GoptcaseA}
\end{align}

\subsection{Optimization for nonidentical oscillators with frequency mismatch}

We next consider the general case with nonidentical oscillators with a frequency
mismatch $\Delta \omega \geq 0$.
We constrain the Frobenius norm of $K$ as $\| K \|^2 = P$ as before, and
also require that the given $\phi^*$ satisfies Eq.~(\ref{const1}), 
i.e., $\Delta \omega + \Gamma_a(\phi^*) = 0$,
so that $\phi^*$ is actually the stationary phase difference of the oscillators.

We thus seek for the optimal coupling matrix $K_{\rm opt}$
that maximizes
\begin{align}
- \Gamma'_a(\phi^*) = - \left. \frac{d}{d \phi} \Gamma_a(\phi)  \right|_{\phi=\phi^*},
\end{align}
now for a given stationary phase difference $\phi^*$, subject to 
\begin{align}
\| K \|^2 = P \quad (P>0)
\end{align}
and
\begin{align}
\Delta \omega + \Gamma_a(\phi^*) = 0.
\label{const1}
\end{align}
Here, we exclude the cases with $\phi^* = 0$ and $\phi^*=\pm \pi$, because these states
 can never be realized when $\Delta \omega > 0$ as we argue later (the case with
$\Delta \omega = 0$ and $\phi^*=0$ was already considered in the previous subsection,
and $\Delta \omega = 0$ and $\phi^*=\pm \pi$ can be analyzed similarly).

Using Lagrange multipliers $\lambda$ and $\mu$, we introduce an action
(in the rest of this subsection, shorthand notations $V_* = V(\phi_*)$ and $V'_* = V'(\phi_*)$ are used),
\begin{align}
S( K, \lambda, \mu )
=& - \left. \frac{d}{d \phi} \Gamma_a(\phi)  \right|_{\phi=\phi^*} + \lambda \Big( \| K \|^2 - P \Big)
\cr
& + \mu \Big( \Delta \omega + \Gamma_a(\phi^*) \Big)
\cr
=& - {\rm Tr\ } \Big( K V_*^{'T} \Big) + \lambda \Big( \| K \|^2 - P \Big) 
\cr
& + \mu \Big( \Delta \omega + {\rm Tr\ } ( K V_*^{\rm T} ) \Big).
\end{align}
Differentiating $S$ by $K$, $\lambda$, and $\mu$, we obtain 
\begin{align}
\frac{\partial}{\partial K} S( K, \lambda, \mu )
=& 
- V'_*+2 \lambda K + \mu V_*
= 0
\end{align}
and the two constraints, Eqs.~(\ref{const1}) and (\ref{const0}). Thus, the optimal $K$ should satisfy
\begin{align}
K = \frac{1}{2 \lambda} ( V'_* - \mu V_* ),
\end{align}
and plugging this into Eq.~(\ref{const1}) yields
\begin{align}
\Delta \omega + \frac{1}{2\lambda} {\rm Tr\ } ( V'_* V_*^{\rm T} ) - \frac{\mu}{2\lambda} \| V_* \|^2 = 0.
\end{align}
Solving this equation for $\mu$, we obtain
\begin{align}
\mu= \frac{2 \lambda \Delta \omega + {\rm Tr\ } ( V'_* V_*^{\rm T} ) }{ \| V_* \|^2}
\end{align}
and therefore
\begin{align}
K_{\rm opt} = \frac{1}{2 \lambda} \left( V'_* - \frac{ 2 \lambda \Delta \omega + {\rm Tr\ } ( V'_* V_*^{\rm T} ) }{ \| V_* \|^2 } V_*\right),
\label{koptcaseC}
\end{align}
where $\lambda$ has yet to be determined from the constraint on the Frobenius norm.

Plugging this $K_{\rm opt}$ into $\| K \|^2 = P$ and using ${\rm Tr\ } (V' V^{\rm T}) = {\rm Tr\ } (V V'^{\rm T})$, we obtain
\begin{align}
4 \Big( (\Delta \omega)^2 - {\|V_* \|^2} P \Big) \lambda^2 + \| V'_* \|^2 {\|V_*\|^2} - [ {\rm Tr\ }(V'_* V_*^{\rm T}) ]^2 = 0,
\end{align}
which gives
\begin{align}
\lambda = \pm \frac{1}{2} \sqrt{ \frac{ \| V'_* \|^2 \| V_* \|^2 - [ {\rm Tr\ } ( V'_* V_*^{\rm T} ) ]^2 }{ \| V_* \|^2 P - ( \Delta \omega )^2 }}.
\label{lambdacase3}
\end{align}
It turns out that the minus sign should be chosen to maximize the linear stability (see below).

Note here that $\| V'_* \| \| V_* \| \geq {\rm Tr\ } ( V'_* V_*^{\rm T} )$ holds by the Schwartz inequality (see Appendix A), so the condition
\begin{align}
P > \frac{ ( \Delta \omega )^2 }{ \| V_* \|^2 }
\label{necessary1}
\end{align}
is necessary for $\lambda$ and hence $K_{\rm opt}$ to exist.
Note also that 
\begin{align}
\| V'_* \| \| V_* \| > {\rm Tr\ } ( V'_* V_*^{\rm T} )
\end{align}
should hold strictly for the existence of $K_{\rm opt}$ in Eq.~(\ref{koptcaseC}), that is, $\| V'_* \| \| V_* \|$ should not be equal to ${\rm Tr\ } ( V'_* V_*^{\rm T} )$, because then $\lambda = 0$ and $K_{\rm opt}$ does not exist.
Therefore, the optimization problem cannot be solved in the case that $V'_*$ and $V_*$ are parallel to each other.

The antisymmetric part of the phase coupling function for $K_{\rm opt}$ is given by
\begin{align}
\Gamma_a(\phi)
=&
{\rm Tr\ }\{ K_{\rm opt} V(\phi)^{\rm T} \}
\cr
=&
\frac{1}{2\lambda} {\rm Tr\ } \{ V_*' V(\phi)^{\rm T} \} 
- 
\frac{1}{2\lambda} \frac{ {\rm Tr\ } ( V_*' V_*^{\rm T} ) }{\| V_* \|^2} {\rm Tr\ } \{ V_* V(\phi)^{\rm T} \} 
\cr
&-
\frac{ \Delta \omega }{ \| V_* \|^2 } {\rm Tr\ } \{ V_* V(\phi)^{\rm T} \}
\end{align}
and the maximal possible linear stability is given by
\begin{align}
- \Gamma_a'(\phi^*)
=& -  {\rm Tr\ } ( K_{\rm opt} V_*^{'T} ) \cr
=& - \frac{1}{2 \lambda \| V_* \|^2} \left\{ \| V'_* \|^2 \| V_* \|^2 -  [ {\rm Tr\ } ( V'_* V_*^{\rm T} ) ]^2 \right\}
\cr
&+  \frac{ \Delta \omega }{ \| V_* \|^2} {\rm Tr\ } ( V'_* V_*^{\rm T} ).
\label{casecmaxstab}
\end{align}
Because $\| V_* \|^2 > 0$ and $\| V'_* \|^2 \| V_* \|^2 - [ {\rm Tr\ } ( V'_* V_*^{\rm T} ) ]^2 \geq 0$, the first term is positive only when $\lambda < 0$.
Therefore, the minus sign should be chosen for $\lambda$ in Eq.~(\ref{lambdacase3}) to realize the maximal stability, and the optimal coupling matrix is given by Eq.~(\ref{koptcaseC}) with the negative $\lambda$.

Note that even if we choose the minus sign for $\lambda$, the above quantity can still be negative if the second term on the right-hand side is negative, i.e., ${\rm Tr\ } ( V'_* V_*^{\rm T} ) < 0$. If so, the fixed point with phase difference $\phi^*$ is unstable and cannot be realized.
Thus, in this case, as can be shown by comparing the two terms on the right-hand side of Eq.~(\ref{casecmaxstab}), $P$ should additionally satisfy
\begin{align}
	P > \frac{ ( \Delta \omega )^2 }{ \| V_* \|^2 - [ {\rm Tr\ } ( V'_* V_*^{\rm T} ) ]^2 / \| V'_* \|^2 },
	\label{necessary2}
\end{align}
for $\phi^*$ to be linearly stable.

Depending on the conditions, the present optimization problem may or may not possess an appropriate solution.
For example, when $\Delta \omega \neq 0$, it is impossible to realize 
completely in-phase ($\phi^*=0$) or anti-phase ($\phi^*= \pm \pi$) synchronization, because 
$\Gamma_a(\phi)$ satisfies $\Gamma_a(0) = \Gamma_a(\pm \pi) = 0$, so $\Delta \omega + \Gamma_a(\phi) = 0$
can never be satisfied at $\phi=0$ or $\phi= \pm \pi$.
Also, when $\Delta \omega \neq 0$, it is generally difficult (very large $P$ is required) to realize the synchronized state with a stationary phase difference $\phi^*$ very close to $0$ or $\pi$. This will be illustrated in the next section.
The equation $\Delta \omega + \Gamma(\phi) = 0$ may also have multiple solutions,
so not only the fixed point with the given phase difference but also spurious fixed points with other phase differences may arise.

\begin{figure}[htbp]
\centering
\includegraphics[width=0.8\hsize]{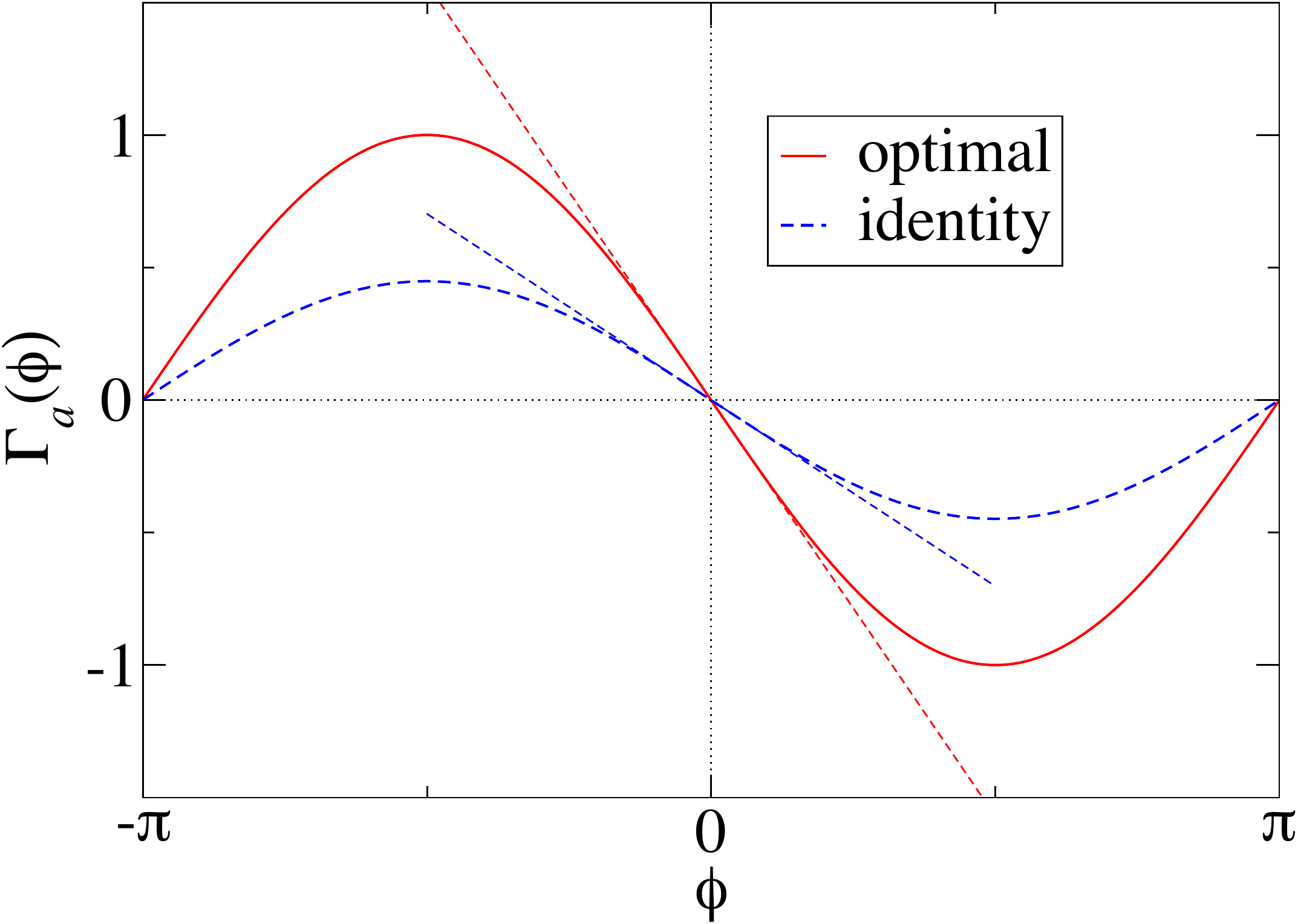}
\caption{Antisymmetric part $\Gamma_a(\phi)$ of the phase coupling function of coupled Stuart-Landau oscillators. The cases with identity coupling and optimal coupling are compared for $P=0.1$. Straight lines represent the slopes at the origin.}
\label{fig01}
\end{figure}

\begin{figure}[htbp]
\centering
\includegraphics[width=0.8\hsize]{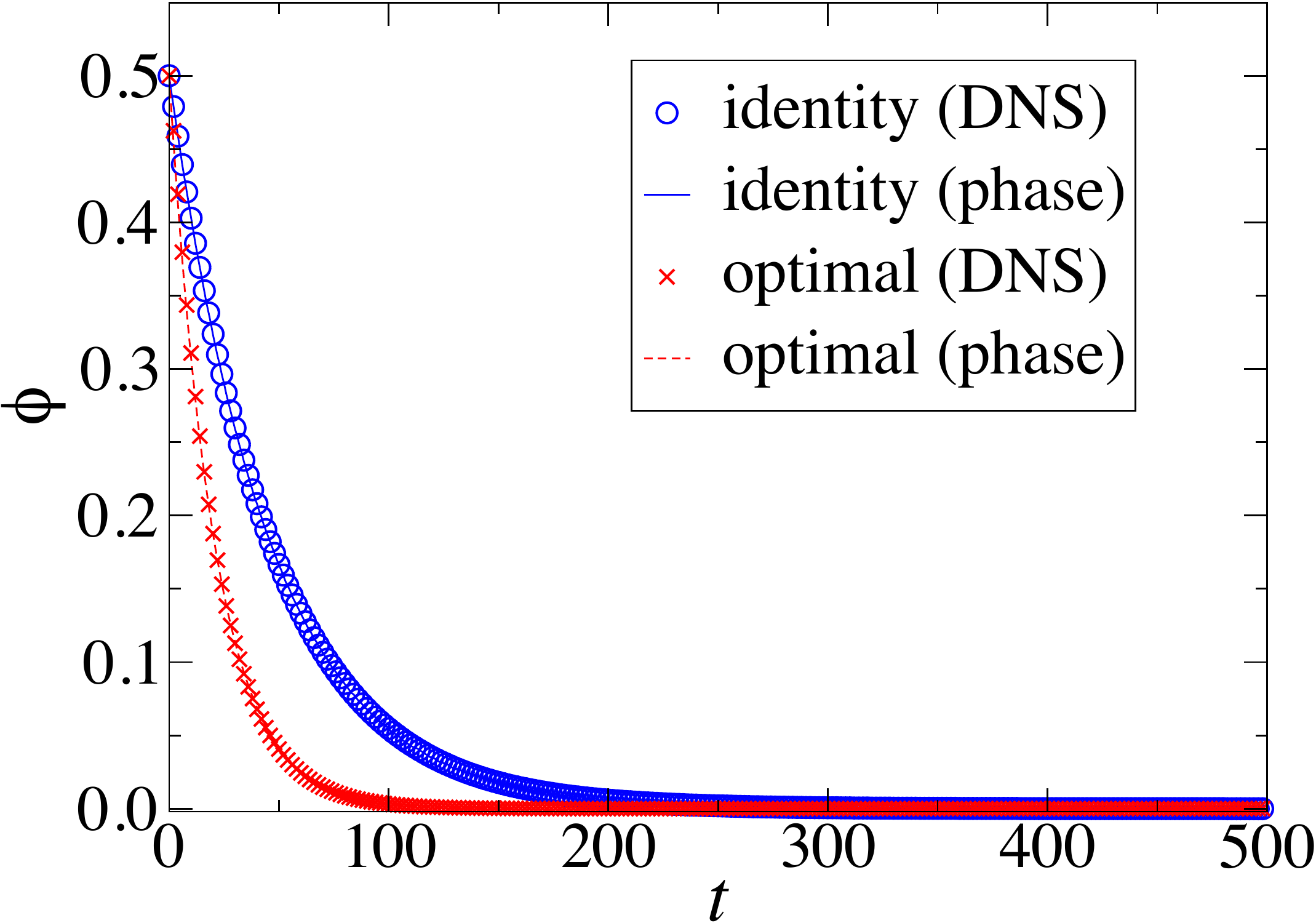}
\caption{In-phase synchronization dynamics of coupled identical Stuart-Landau oscillators. The cases with identity coupling and optimal coupling are compared for $P=0.1$. Results obtained  by direct numerical simulations of the coupled Stuart-Landau oscillators and by numerically integrating reduced phase equations are shown. }
\label{fig02}
\end{figure}

\section{Examples}

In this section, we illustrate the improvement in the linear stability of coupled oscillators
by optimizing $K$ with a few types of limit-cycle oscillators as examples.

\subsection{Stuart-Landau oscillator}

As the first example, we consider the Stuart-Landau (SL) oscillator, a normal form of the supercritical Hopf bifurcation~\cite{Kuramoto}.
All necessary quantities can be analytically calculated for this model.
The SL oscillator has a two-dimensional state variable, ${\bm X} = (x, y)^{\rm T}$, whose dynamics is specified by a vector field
\begin{align}
{\bm F}({\bm X}) =
\left( \begin{array}{c} F_x(x, y) \\ F_y(x, y) \end{array} \right) = 
\left( \begin{array}{c} x - \alpha y - ( x - \beta y) (x^2+y^2) \\ \alpha x + y - ( \beta x + y ) ( x^2+y^2 )  \end{array} \right),
\end{align}
where $\alpha$ and $\beta$ are parameters.
It possesses a single stable limit-cycle orbit of frequency $\omega = \alpha - \beta$ given by
\begin{align}
{\bm X}_0(\theta)
= \left( \begin{array}{c} x_0(\theta) \\ y_0(\theta) \end{array} \right)
= \left( \begin{array}{c} \cos \theta \\ \sin \theta \end{array} \right)
\label{sllc}
\end{align}
with $\theta(t) = \omega t$ (mod $2\pi$).
The phase sensitivity function of this limit cycle can be explicitly calculated as~\cite{Kuramoto,ref:nakao16}
\begin{align}
{\bm Z}(\theta)
= \left( \begin{array}{c} Z_x(\theta) \\ Z_y(\theta) \end{array} \right) 
= \left( \begin{array}{c} - \sin \theta - \beta \cos \theta \\  \cos \theta - \beta \sin \theta \end{array} \right).
\label{slZ}
\end{align}

We consider a pair of symmetrically coupled SL oscillators with identical properties obeying Eq.~(\ref{model}), which is explicitly described by
\begin{align}
\left( \begin{array}{c} \dot{x}_1 \\ \dot{y}_1 \end{array} \right)
=&
\left(\begin{array}{c} F_x(x_1, y_1) \\ F_y(x_1, y_1) \end{array}\right)
+
\epsilon
\left( \begin{array}{cc} K_{11} & K_{12} \\ K_{21} & K_{22} \end{array} \right)
\left( \begin{array}{c} x_2 - x_1 \\ y_2 - y_1 \end{array} \right),
\cr
\left( \begin{array}{c} \dot{x}_2 \\ \dot{y}_2 \end{array} \right)
=&
\left(\begin{array}{c} F_x(x_2, y_2) \\ F_y(x_2, y_2) \end{array}\right)
+
\epsilon
\left( \begin{array}{cc} K_{11} & K_{12} \\ K_{21} & K_{22} \end{array} \right)
\left( \begin{array}{c} x_1 - x_2 \\ y_1 - y_2 \end{array} \right),
\label{coupledSL}
\end{align}
where ${\bm X}_1=(x_1, y_1)^{\rm T}$ and ${\bm X}_2 = (x_2, y_2)^{\rm T}$ are the state variables of the oscillators.
The frequency of the oscillators in the absence of mutual coupling is given by $\omega = \alpha - \beta$.

From Eqs.~(\ref{sllc}) and (\ref{slZ}), the matrix $V(\phi)$ and its derivative $V'(\phi)$ can be calculated as
\begin{align}
V(\phi)
&=
-\sin \phi \left( \begin{array}{cc}
1 & -\beta \\
\beta & 1
\end{array} \right),
\label{glV}
\end{align}
and
\begin{align}
V'(\phi)
&=
-\cos \phi \left( \begin{array}{cc}
1 & - \beta \\
\beta & 1
\end{array} \right),
\label{gldV}
\end{align}
so they are parallel to each other.

Because the oscillators are identical, $\omega_1 = \omega_2$ and $\Delta \omega = 0$,
the in-phase synchronized state $\phi^* = 0$ always exists. 
From Eq.~(\ref{KoptcaseA}), the stability of this state can be maximized by choosing $K$ as
\begin{align}
K_{\rm opt} = \sqrt{\frac{P}{2 (\beta^2+1)}}\left( \begin{array}{cc}
1 & - \beta \\
\beta & 1
\end{array} \right),
\label{glkopt}
\end{align}
and the maximum possible linear stability is given, from Eq.~(\ref{GoptcaseA}), by
\begin{align}
- \Gamma_a'(0) = \sqrt{P} \sqrt{2 (\beta^2+1)}.
\end{align}

For comparison, suppose that the coupling matrix is a multiple of
the identity matrix with the same Frobenius norm as $K_{\rm opt}$, i.e., 
\begin{align}
K_I = \sqrt{\frac{P}{2}} \left( \begin{array}{cc}
1 & 0 \\
0 & 1
\end{array} \right),
\label{identity}
\end{align}
which we call the ``identity coupling'' hereafter.
The linear stability of $\phi^*=0$ with this $K_I$ is given by
\begin{align}
- \Gamma_a'(0)
=
- {\rm Tr\ } ( K_I V'(0)^{\rm T} )
=  \sqrt{2 P},
\end{align}
so the linear stability improves by a factor of $\sqrt{\beta^2+1}$ by using the optimal coupling matrix $K_{\rm opt}$ from the case with $K_I$.

\begin{figure}[htbp]
\centering
\includegraphics[width=0.8\hsize]{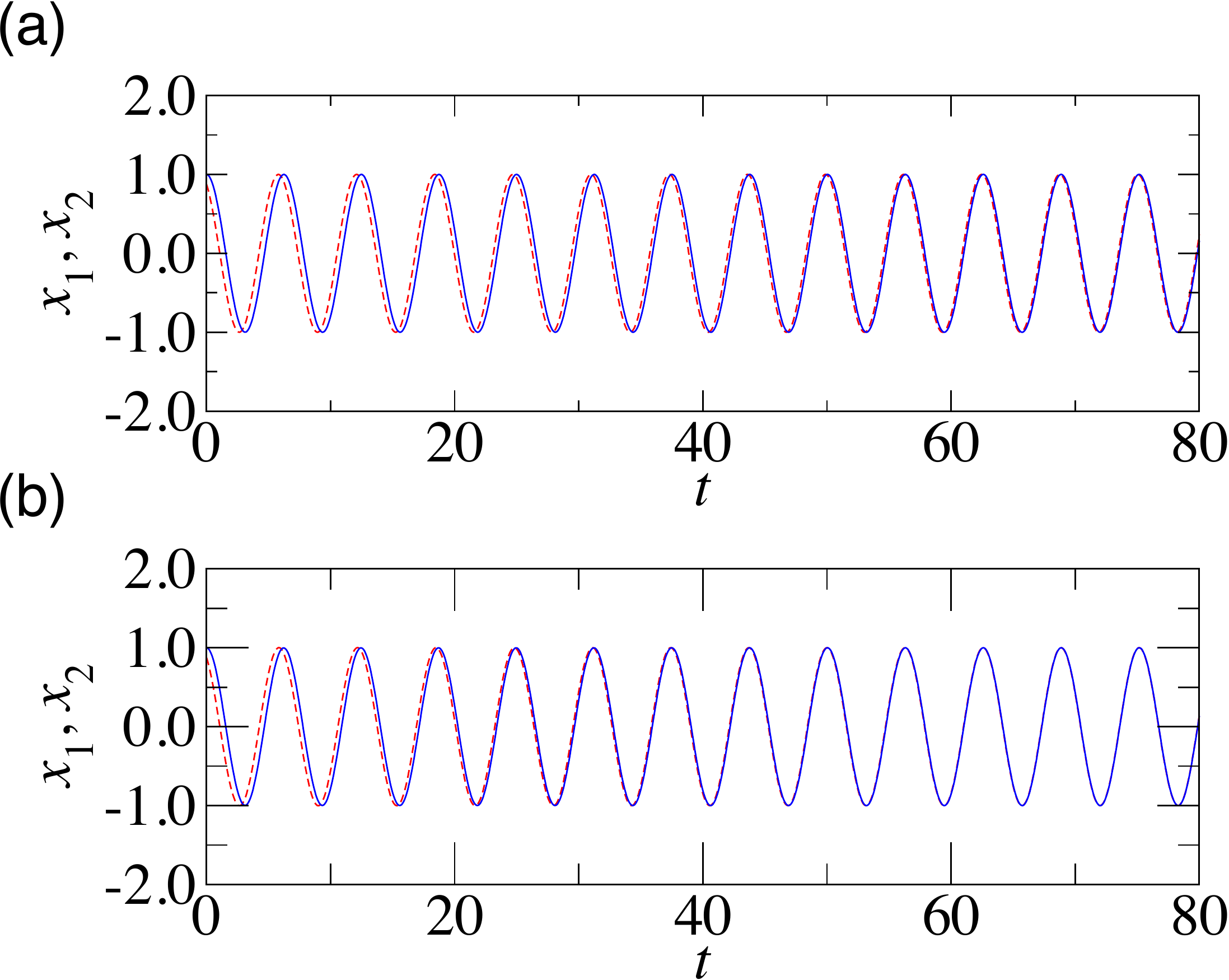}
\caption{In-phase synchronization dynamics of coupled Stuart-Landau oscillators for $P=0.1$ and $\epsilon=0.05$. Identity coupling (a) vs. optimal coupling (b).}
\label{fig03}
\end{figure}

Figure~\ref{fig01} shows the antisymmetric part $\Gamma_a(\phi)$ of the phase coupling function  calculated for coupling matrices
$K = K_{\rm opt}$ and $K = K_I$, with parameters $\alpha=3$ and $\beta=2$ and overall coupling intensity $P=0.1$.
It can be seen that the linear stability $-\Gamma_a'(0)$ of the in-phase fixed point $\phi^* = 0$ is higher
in the case with $K_{\rm opt}$ than in the case with $K_I$.
Figure~\ref{fig02} compares the time courses of the phase difference $\phi(t)$ from the initial condition $\phi(0)=0.5$ for $K_{\rm opt}$ and $K_I$
obtained by direct numerical simulations of the coupled SL oscillators and by numerical integration of the reduced phase equation.
Figure~\ref{fig03} shows the synchronization dynamics obtained numerically, where the $x$ components of the two oscillators are shown
for $K_{\rm opt}$ and $K_I$.
It can be seen that the oscillators synchronize faster in the case with $K_{\rm opt}$, reflecting higher linear stability of $\phi^* = 0$, than in the case with $K_I$.

It is interesting to note that when $\beta = 0$, that is, when the instantaneous frequency of the
SL oscillator does not depend on its amplitude, $K_I$ is already optimal and no improvement can be made
by introducing cross coupling between different variables of the oscillators.

For nonidentical SL oscillators with a small parameter mismatch $\Delta \omega > 0$, the antisymmetric part of the phase coupling function and its derivative take the form $\Gamma_a(\phi) = - C \sin \phi$ and $\Gamma_a'(\phi) = - C \cos \phi$ from Eqs.~(\ref{Kphasecoupling}), (\ref{glV}) and (\ref{gldV}), where $C$ is a constant determined by $K$ and $\beta$ ($K$ should also satisfy $\| K \|^2 = P$). Once the stationary phase difference $\phi^* \neq 0$ is specified, the constant $C$ is determined as $C = \Delta \omega / \sin \phi^*$ and the linear stability of $\phi^*$ is given by a fixed value $-\Gamma_a'(\phi^*) = \Delta \omega / \tan \phi^*$. Therefore, we cannot consider further optimization of the stability for the coupled SL oscillators.
Indeed, we cannot consider the second case in Sec. III, because $V(\phi)$ and $V'(\phi)$,
given by Eqs.~(\ref{glV}) and (\ref{gldV}), are strictly parallel to each other,
so the Lagrange multiplier $\lambda$ vanishes and $K_{\rm opt}$ does not exist.
This is a peculiar property of the SL oscillator with purely sinusoidal
limit cycles and phase sensitivity functions.

\begin{figure}[htbp]
\centering
\includegraphics[width=0.8\hsize,clip]{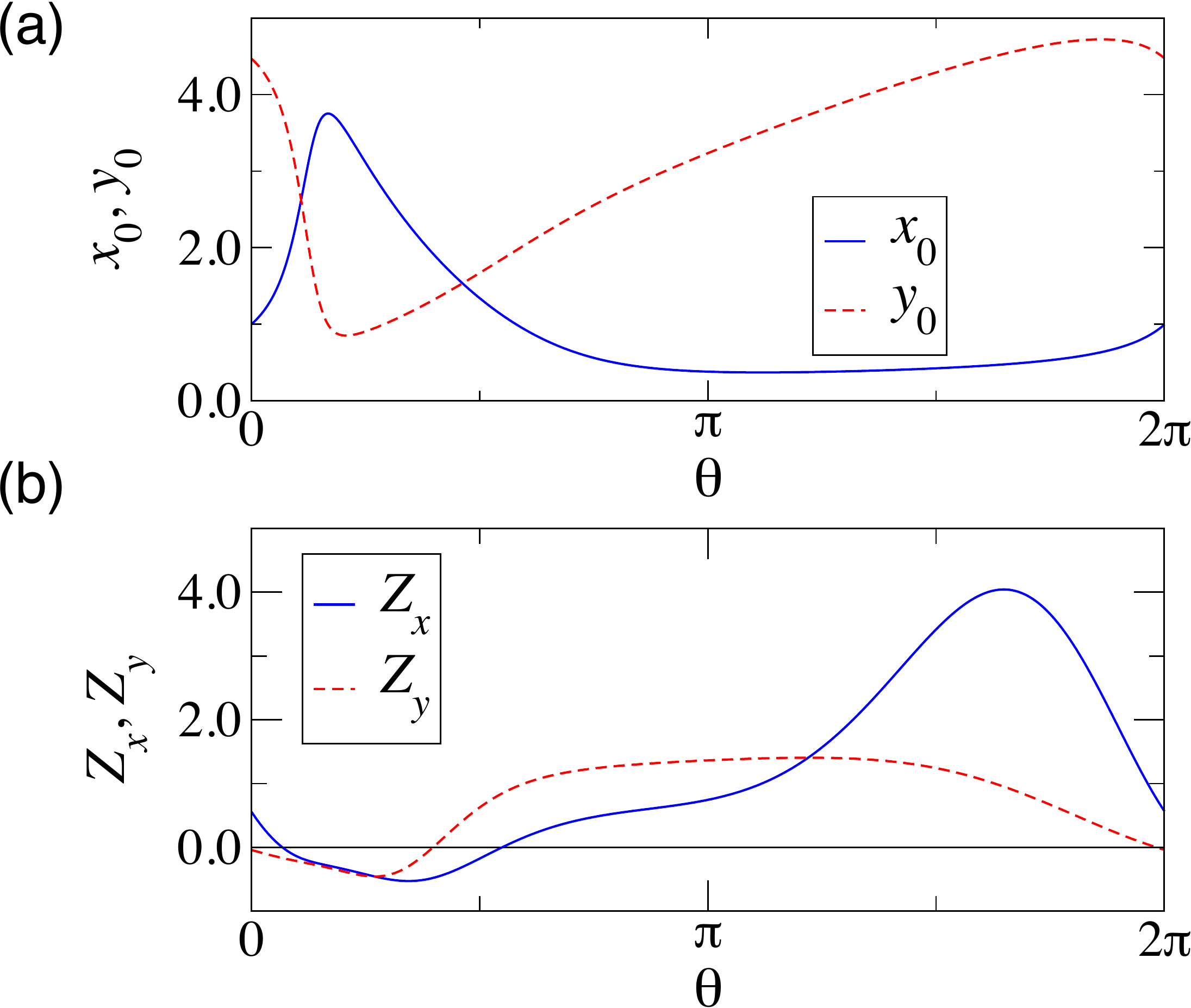}
\caption{Limit-cycle solution ${\bm X}_0(\theta) = (x_0(\theta), y_0(\theta))^{\rm T}$ (a)
and phase sensitivity function ${\bm Z}(\theta) = (Z_x(\theta), Z_y(\theta))^{\rm T}$ (b) of the Brusselator.}
\label{fig04}
\end{figure}

\subsection{Brusselator}

As the second example, we use the Brusselator model of chemical oscillations~\cite{Kuramoto}.
It has a two-dimensional state variable, ${\bm X} = (x, y)^{\rm T}$, which obeys
\begin{align}
{\bm F}({\bm X}) = \left( \begin{array}{c}
 a - ( b + 1 ) x + x^2 y \\ b x - x^2 y
\end{array} \right),
\end{align}
where $a$ and $b$ are parameters.
When $a = 1$ and $b = 3$, the period of the oscillation is $\omega \approx 0.878$.
Figure~\ref{fig04} shows the limit-cycle solution ${\bm X}_0(\theta) = (x_0(\theta), y_0(\theta))^{\rm T}$ and the phase sensitivity function ${\bm Z}(\theta) = (Z_x(\theta), Z_y(\theta))^{\rm T}$ for $0 \leq \theta < 2\pi$ obtained numerically.
Other quantities such as $V(\phi)$ and $V'(\phi)$ can also be numerically calculated from these ${\bm X}_0(\theta)$ and ${\bm Z}(\theta)$.

We consider a pair of Brusselators with parameters $b = 3 \pm \delta$,
where $\delta$ is a small number representing parameter mismatch, and couple them
in the same way as in the previous SL case, Eq.~(\ref{coupledSL}).
We seek for the optimal $K_{\rm opt}$ that gives the maximum stability of the
synchronized state, and compare the results for $K_{\rm opt}$ 
with those for the identity coupling, i.e., $K_I$ given by Eq.~(\ref{identity}).
The overall intensity $P$ is fixed at $P = 0.1$ in the following.

\begin{figure}[htbp]
\centering
\includegraphics[width=0.8\hsize,clip]{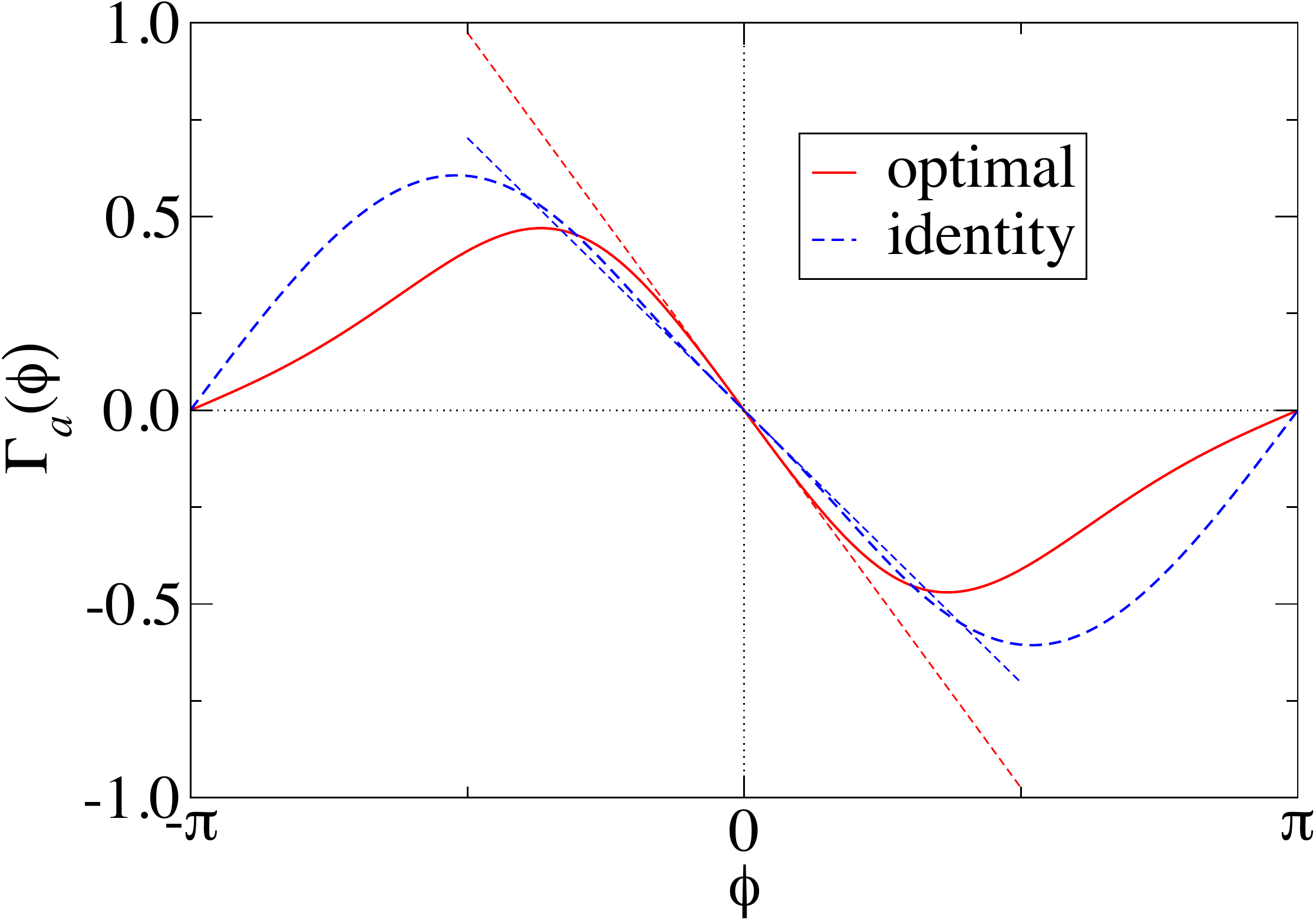}
\caption{Antisymmetric part  $\Gamma_a(\phi)$ of the phase coupling function of coupled Brusselators. The results for optimal and identity coupling matrices $K_{\rm opt}$ and $K_I$ are compared for $P=0.1$. Straight lines represent the slopes at the origin.}
\label{fig05}
\end{figure}

We first consider the case without parameter mismatch, $\delta = 0$.
The frequencies of the oscillator are identical in this case, $\Delta \omega = 0$.
The optimal and identity coupling matrices with $\| K \|^2 = P = 0.1$ are calculated as
\begin{align}
K_{\rm opt} \approx \left(\begin{array}{cc}
0.0972 & 0.195 \cr
-0.0428 & 0.225 \cr
\end{array} \right),
\quad
K_I \approx \left(\begin{array}{cc}
0.224 & 0 \cr
0 & 0.224
\end{array} \right).
\end{align} 
We can see that, in the optimal case, the feedback from the difference in $y$ component to the dynamics of $x$ and $y$ components is stronger than that in the opposite direction. This reflects the waveforms of the oscillation and phase sensitivity function, in particular, that the variation in $y$ is generally larger than that in $x$, as shown in Fig.~\ref{fig04}.

\begin{figure}[htbp]
\centering
\includegraphics[width=0.8\hsize]{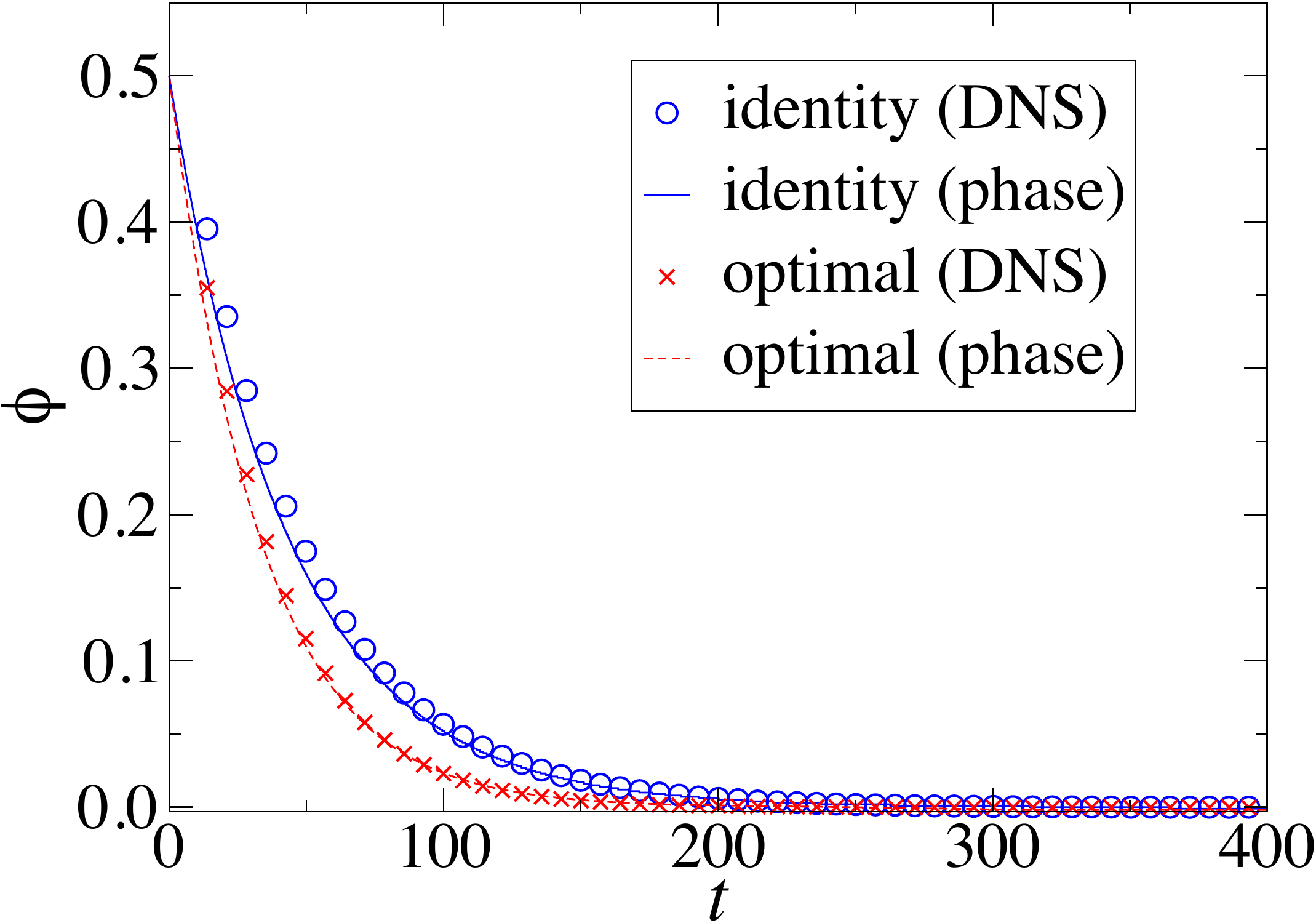}
\caption{Evolution of phase difference $\phi$ for identity and optimal coupling matrices, $K_I$ and $K_{\rm opt}$, starting from initial phase difference $\phi = 0.5$. Results obtained by direct numerical simulations (DNS) of coupled Brusselators and by numerically integrating the reduced phase equations are shown. The parameters are $P=0.1$ and $\epsilon = 0.05$.}
\label{fig06}
\end{figure}

\begin{figure}[htbp]
\centering
\includegraphics[width=0.8\hsize,clip]{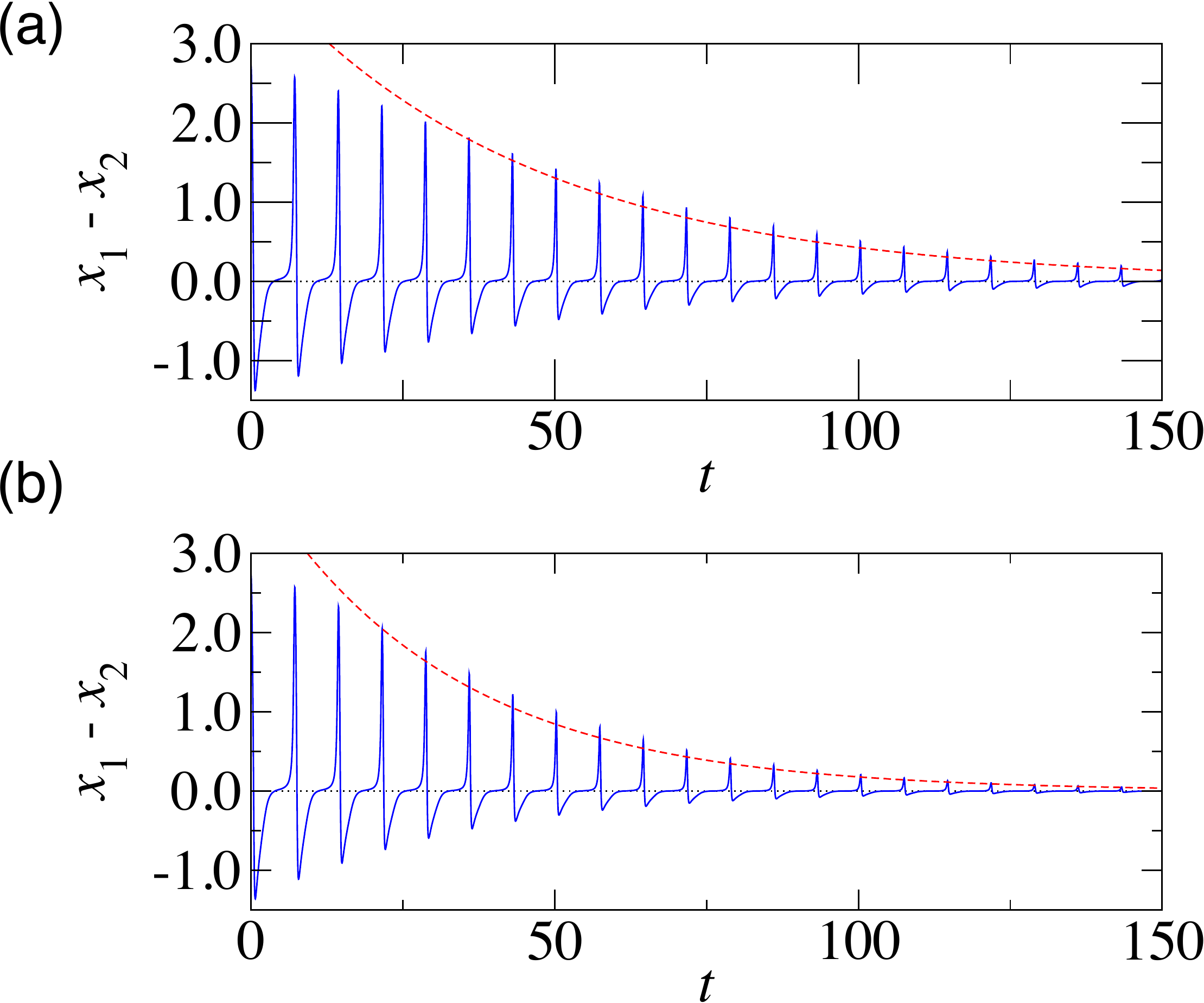}
\caption{In-phase synchronization process of two identical Brusselators with identity (a) and optimal (b) coupling matrices, $K_{\rm opt}$ and $K_I$. In each figure, time sequence of the difference in $x$ components between the two oscillators is plotted by a solid line, and the dashed line indicates an exponentially decaying curve with a decay rate $\epsilon \Gamma_a'(0) < 0$ (actually $4 \exp [ \epsilon \Gamma'_a(0) t ]$) . The parameters are $P=0.1$ and $\epsilon = 0.05$.}
\label{fig07}
\end{figure}

Figure~\ref{fig05} shows the antisymmetric part $\Gamma_a(\phi)$ of the phase coupling function for
$K=K_{\rm opt}$ and $K=K_I$.
The linear stability of the in-phase synchronize state $\phi^* = 0$ is approximately $-\Gamma_a'(0)=0.621$ for $K_{\rm opt}$ and $-\Gamma_a'(0)=0.448$ for $K_I$.
Figure~\ref{fig06} shows the evolution of phase differences for $K_{\rm opt}$ and $K_I$
obtained by direct numerical simulations of the coupled Brusselators and by
numerical integration of the reduced phase model.
The parameter $\epsilon$ is fixed at $\epsilon = 0.05$ in the numerical simulations.
Figure~\ref{fig07} shows the synchronization processes of the Brusselators,
where time courses of the differences in $x$ components between the oscillators,
i.e., $x_1 - x_2$, are plotted for $K_{\rm opt}$ and $K_I$.
For comparison, an exponentially decaying curve with the decay rate $\epsilon \Gamma_a'(0) \ (<0)$
is also shown in each figure.
It can been seen that in-phase synchronization is established faster when $K_{\rm opt}$ is used, and the exponential decay rate of the state difference matches with the linear stability $\Gamma'_a(0)$ of $\phi^*=0$.

We next consider the case with parameter mismatch, $\delta = 0.01$.
The frequencies of the oscillators are
$\omega_1 \approx 0.8797$ ($b=2.99$) and $\omega_2 \approx 0.8762$
($b=3.01$).
We assume $\epsilon = 0.02$ in the following calculations, so the frequency mismatch parameter is
$\Delta \omega \approx 0.175$.
Using the results obtained in the previous section, we calculate the optimal coupling matrix
$K_{\rm opt}$ for a given phase difference $\phi^*$ in $(-\pi, \pi)$.

\begin{figure}[htbp]
\centering
\includegraphics[width=0.8\hsize,clip]{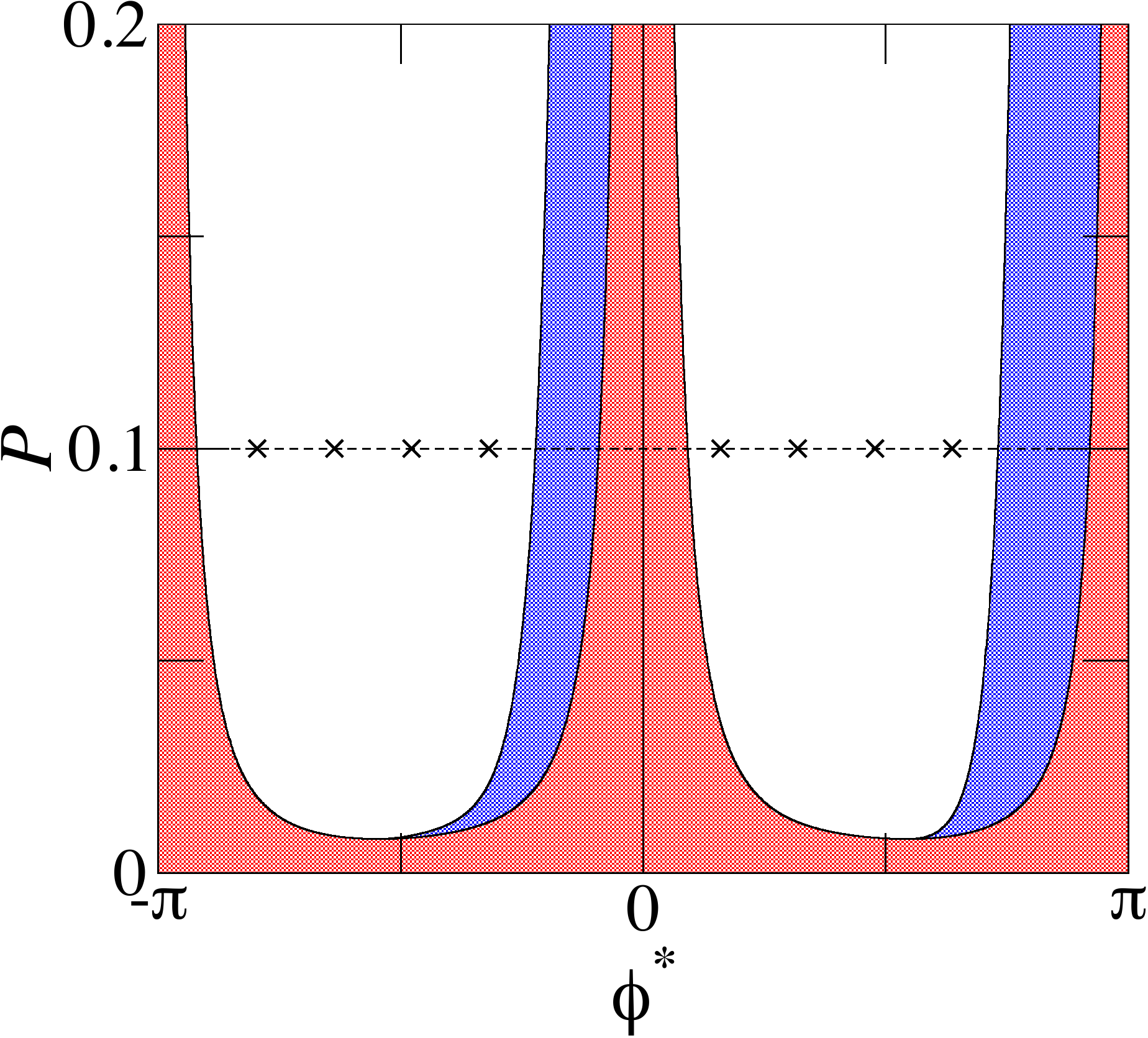}
\caption{Necessary conditions for the overall coupling intensity $P$ plotted as functions of the
target phase difference $\phi^*$ for frequency mismatch $\Delta \omega = 0.175$. Only non-shaded regions are realizable.
The first condition (\ref{necessary1}) is violated in the red-shaded region, and the second condition (\ref{necessary2}) is violated in blue-shaded region. Crosses represent the values of the given phase difference $\phi^*$ used in the example.}
\label{fig08}
\end{figure}

Figure~\ref{fig08} shows the necessary conditions for $P$ given by
Eq.~(\ref{necessary1}) and Eq.~(\ref{necessary2})
as functions of the phase difference $\phi^*$ for $\Delta \omega = 0.175$, where the latter applies only when ${\rm Tr\ } ( V'_* V_*^{\rm T} ) < 0$.
Both conditions are satisfied in the non-shaded regions.
We see that $P$ should not be too small and that the regions near $\phi^*=0$ and $\phi^*=\pm \pi$ are difficult to realize,
as argued in the previous section.

\begin{figure}[htbp]
\centering
\includegraphics[width=0.8\hsize,clip]{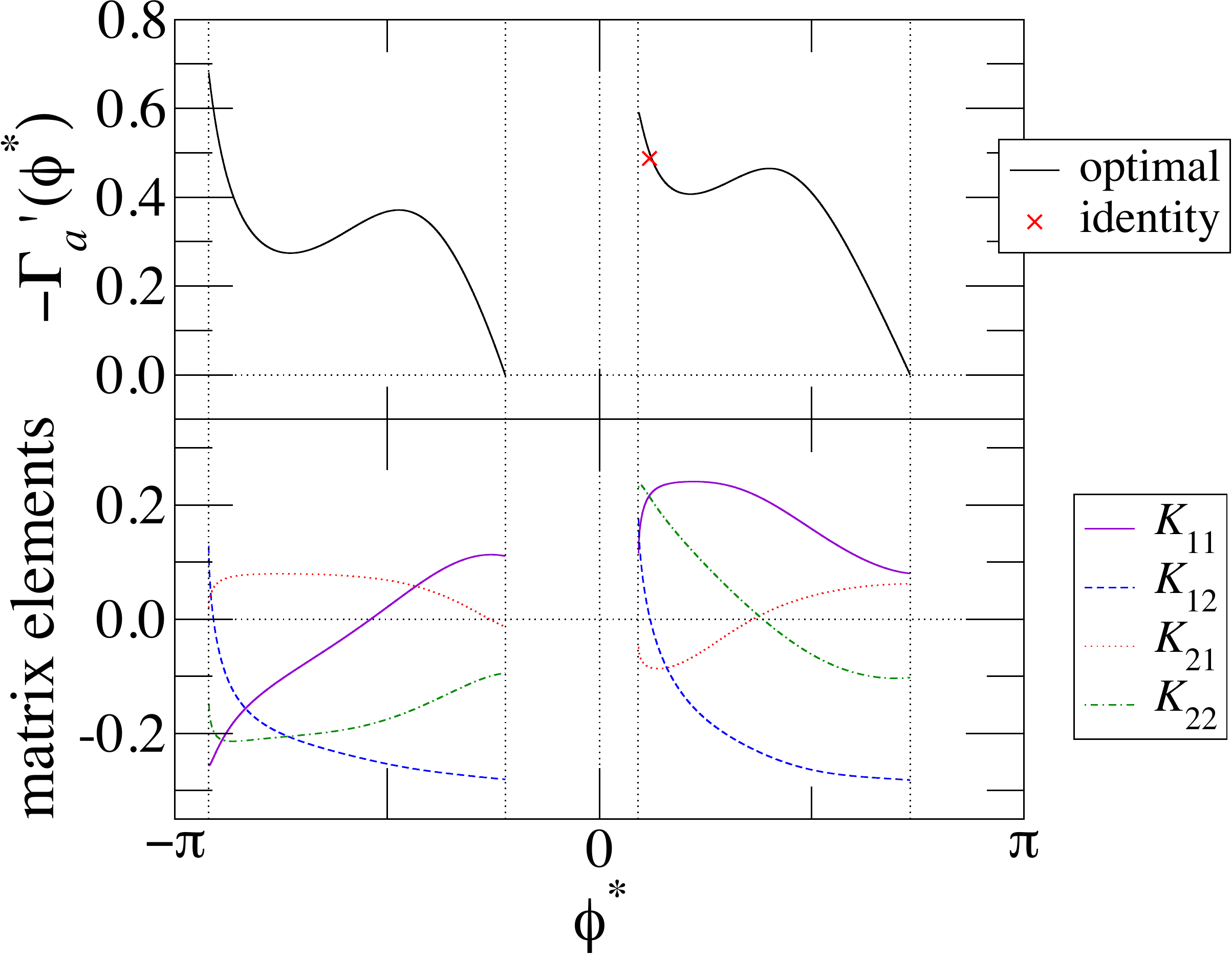}
\caption{Matrix elements of the optimal coupling matrix $K_{11}, K_{12}, K_{21}, K_{22}$
and linear stability $-\Gamma_a'(\phi^*)$ plotted as functions of the phase difference $\phi^*$ for $P=0.1$.
Dotted vertical lines represent the boundaries of the regions in which both of the necessary conditions are satisfied.}
\label{fig09}
\end{figure}

Figure~\ref{fig09} shows the elements of the optimal coupling matrix $K_{\rm opt}$ and the corresponding
stability of the fixed point as functions of the phase difference $\phi^*$. For comparison,
the results for $K_I$, which gives a stable phase difference $\phi^* \approx 0.378$ and negative slope $0.487$,
are also indicated in the figure.
In this particular example, $K_I$ yields reasonably high stability close to the negative slope $0.493$ with the optimal coupling matrix $K_{\rm opt}$ at $\phi^* = 0.378$~\cite{footnote2}.
In the blank regions where the data are not shown, any of the necessary conditions is not satisfied.
The stability varies with $\phi^*$ and, in this case, nearly anti-phase
synchronized state yields the highest stability.
Elements of the coupling can be positive or negative depending on $\phi^*$.

\begin{figure}[htbp]
\centering
\includegraphics[width=0.8\hsize,clip]{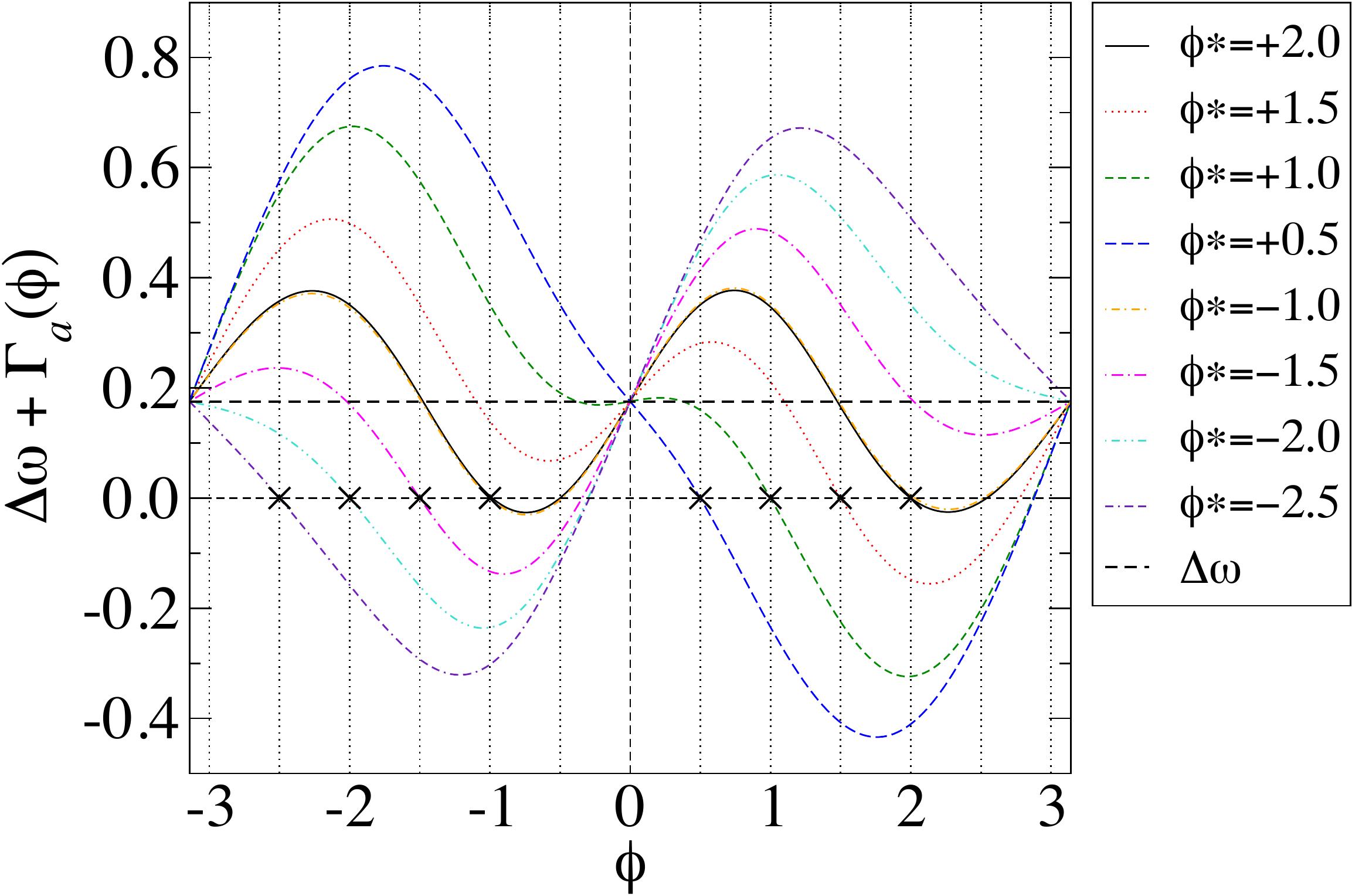}
\caption{Examples of the antisymmetric part of the phase coupling function for given stable phase differences, $\phi^* = 2.0, 1.5, 1.0, 0.5, -1.0, -1.5, -2.0,$ and $-2.5$, for frequency mismatch $\Delta \omega = 0.1754$. Crosses represent stable fixed points.}
\label{fig10}
\end{figure}

\begin{figure}[htbp]
\centering
\includegraphics[width=0.8\hsize,clip]{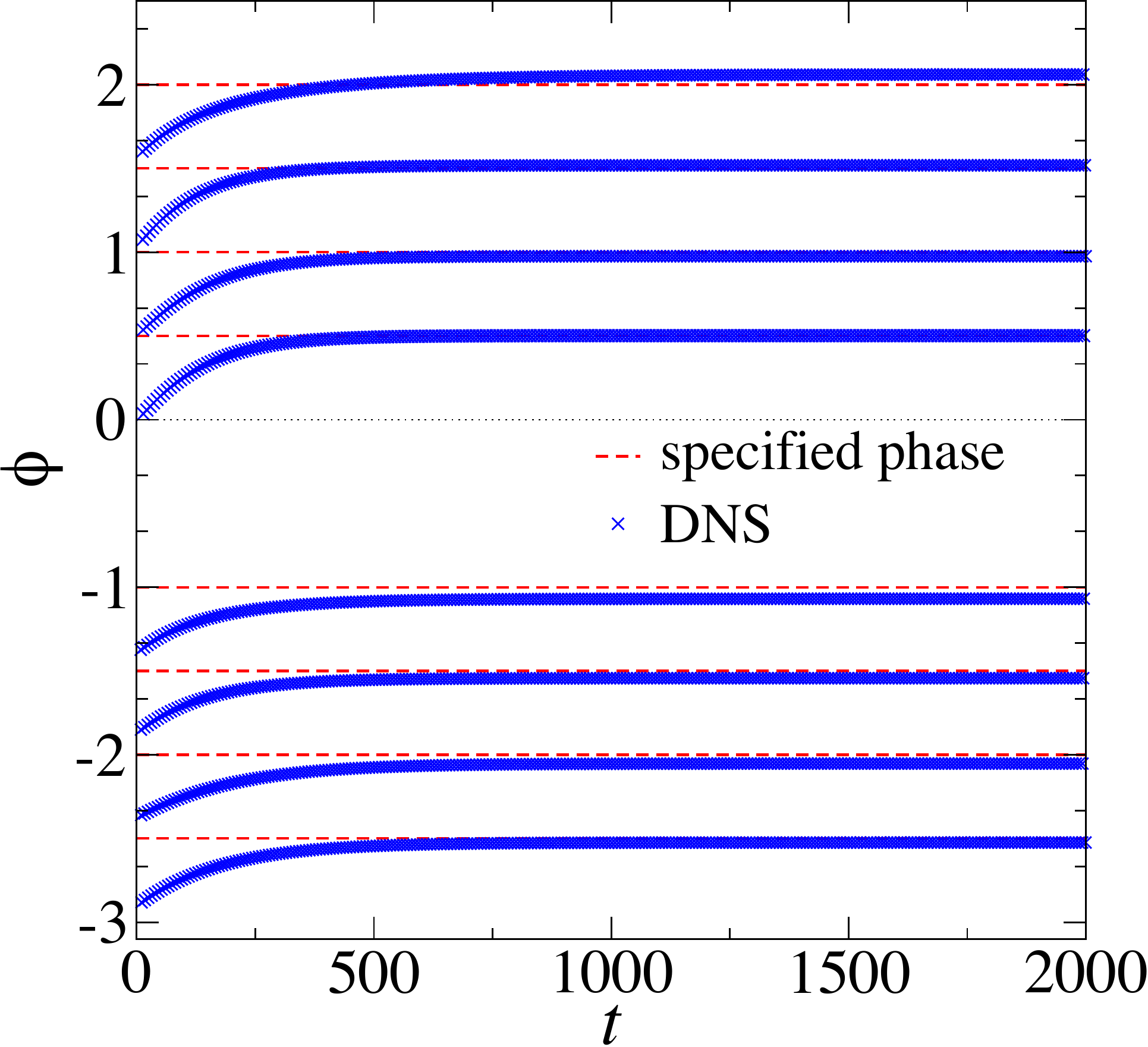}
\caption{Convergence of phase differences to given values ($\phi^* = 2.0, 1.5, 1.0, 0.5, -1.0, -1.5, -2.0,$ and $-2.5$, indicated by horizontal dashed lines). Results of direct numerical simulations are compared with the specified phase values} for frequency mismatch $\Delta \omega = 0.175$ and $\epsilon = 0.02$.
\label{fig11}
\end{figure}

Figure~\ref{fig10} shows the antisymmetric parts $\Gamma_a(\phi)$ of the obtained phase coupling functions for
given phase differences, $\phi^* = -2.5, -2.0, -1.5, -1.0, 0.5, 1.0, 1.5$, and $2.0$. The given phase differences are actually realized with the optimal $\Gamma_a(\phi)$ as stable fixed points. From this figure, we can clearly see why stationary phase differences close to $0$ or $\pi$ are difficult to realize, in consistent with the conditions shown in Fig.~\ref{fig08}.
Figure~\ref{fig11} plots the results of direct numerical simulations for coupled Brusselators using the optimal coupling matrices $K_{\rm opt}$, where the convergence of the phase differences to given values is shown.

\subsection{Lorenz model}

\begin{figure}[htbp]
\centering
\includegraphics[width=0.8\hsize,clip]{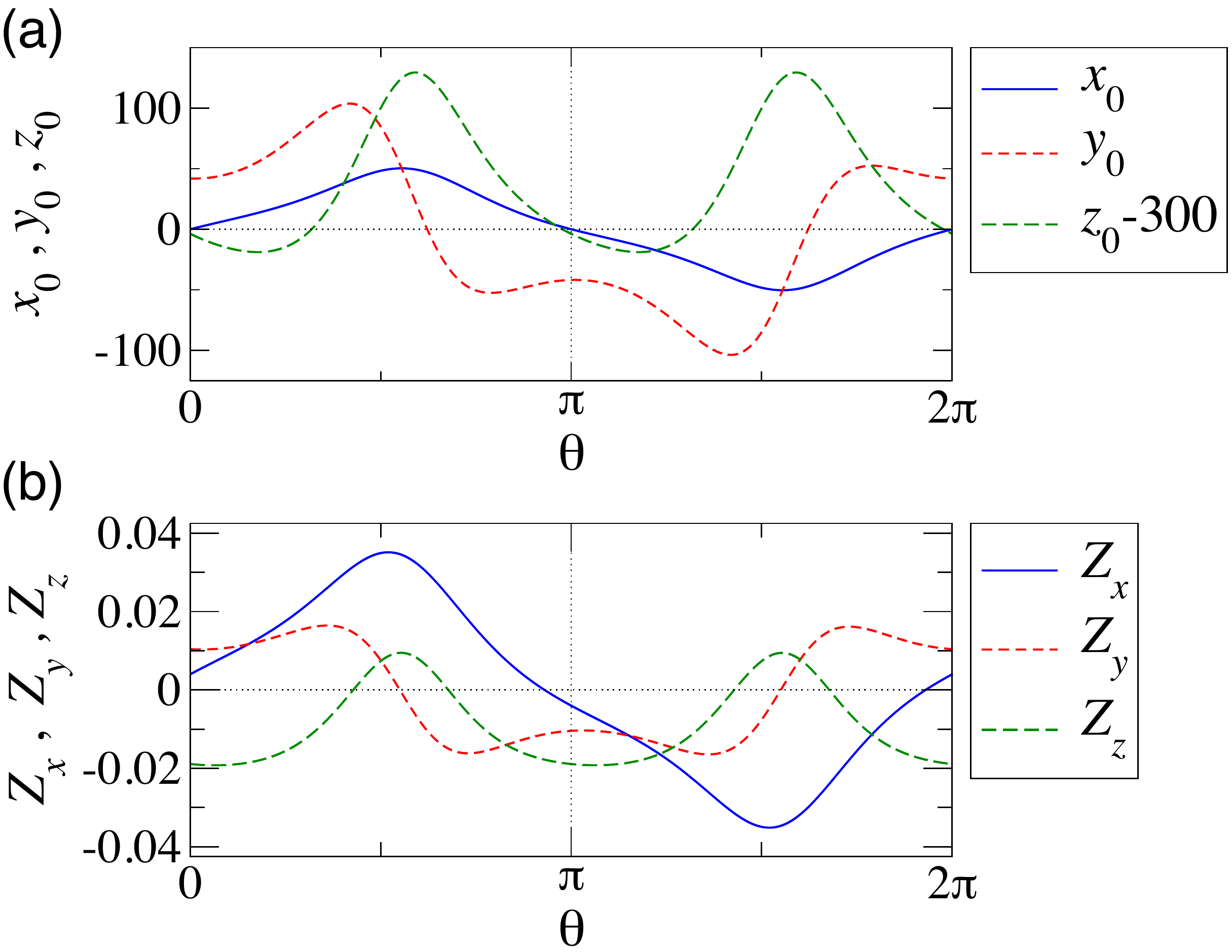}
\caption{(a) Limit-cycle solution ${\bm X}_0(\theta) = (x_0(\theta), y_0(\theta), z_0(\theta))^{\rm T}$  of the Lorenz model, where the $z$ variable is shifted by $300$ for clarity. (b) Phase sensitivity function ${\bm Z}(\theta) = (Z_x(\theta), Z_y(\theta), Z_z(\theta))^{\rm T}$ of the Lorenz model.}
\label{fig12}
\end{figure}

Finally, as a simple three-dimensional example, we consider the Lorenz model in the limit-cycling regime~\cite{strogatz15}, whose state variable ${\bm X} = (x, y, z)^{\rm T}$ evolves with the vector field
\begin{align}
{\bm F}({\bm X}) = \left( \begin{array}{c}
\sigma (y - x) \\
r x - y - x z \\
x y - b z \\
\end{array} \right)
\end{align}
with $\sigma = 10$, $b=8/3$, and $r = 350$.
The frequency of the limit-cycle oscillation is $\omega = 16.18$.
Figure~\ref{fig12} shows the evolution of ${\bm X}_0(\theta) = (x_0, y_0, z_0)^{\rm T}$ for one period of oscillation and the corresponding phase sensitivity function ${\bm Z}(\theta) = (Z_x, Z_y, Z_z)^{\rm T}$ obtained by the adjoint method~\cite{ref:ermentrout10}.

\begin{figure}[htbp]
\centering
\includegraphics[width=0.8\hsize,clip]{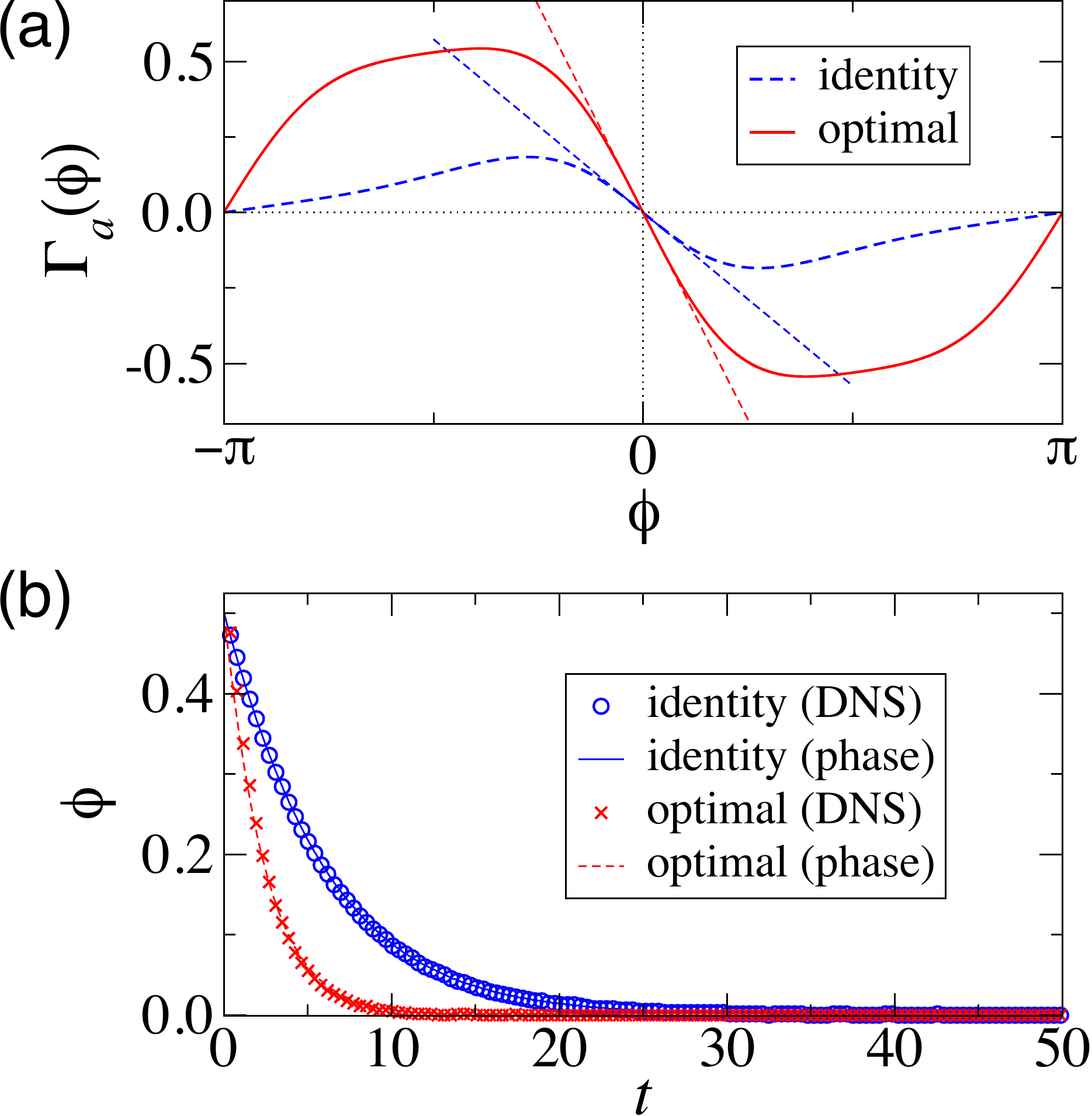}
\caption{(a) Antisymmetric part $\Gamma_a(\phi)$ of the phase coupling function of coupled Lorenz models. Straight line represents the slope at the origin. (b) Convergence of phase difference $\phi(t)$ from $\phi(0) = 0.5$ to $0$. Results of direct numerical simulations and phase models are compared for $\epsilon = 0.5$. In each figure, results for optimal and identity coupling matrices $K_{\rm opt}$ and $K_I$ are compared.}
\label{fig13}
\end{figure}

We consider two Lorenz models without frequency mismatch and couple them via the coupling matrix $K$ as in Eq.~(\ref{model}). 
We compare the results for the optimal coupling matrix $K_{\rm opt}$ with those for $K_I$.
From the results in the previous section, for $P=0.1$, $K_{\rm opt}$ and $K_I$ are estimated as
\begin{align}
K_{\rm opt} \approx \left( \begin{array}{ccc}
0.0283 & -0.263 & 0.000 \\
0.0975 & 0.106 & 0.000 \\
0.000 & 0.000 &  0.095
\end{array} \right)
\end{align}
and
\begin{align}
K_{I} \approx \left( \begin{array}{ccc}
0.183 & 0 & 0 \\
0 & 0.183 & 0 \\
0 & 0 & 0.183
\end{array} \right).
\end{align}
The linear stability $-\Gamma_a'(0)$ is approximately $0.872$ for the optimal coupling and $0.365$ for the identity coupling, respectively.
Figure~\ref{fig13} shows the antisymmetric parts $\Gamma_a(\phi)$ of the phase coupling functions for $K=K_{\rm opt}$ and $K=K_I$, and compares the time courses of the phase differences $\phi$ obtained numerically for $\epsilon  = 0.5$.
We can clearly see that the stability of the in-phase state is higher and correspondingly the phase difference decays to zero faster in the optimal case.

It is notable that $K_{\rm opt}$ has several zero components, indicating no feedback from $z$ component to $x$ or $y$ component nor from $x$ or $y$ component to $z$ component arise even after optimization. This is because $z$ component exhibits qualitatively different dynamics from those of $x$ and $y$ components in the Lorenz model.
As can be seen from Fig.~\ref{fig12}, the fundamental frequency of $z$ component is exactly twice that of $x$ and $y$ components.
Reflecting the symmetry of the Lorenz model (invariance under $x \to -x$, $y \to -y$, $z \to z$), the waveforms of $z_0(\theta)$ and $Z_z(\theta)$ exhibit the same pulse-like oscillations exactly twice while other quantities, $x_0(\theta)$, $y_0(\theta)$, $Z_x(\theta)$, and $Z_y(\theta)$, undergo one period of smooth oscillation that is symmetric to $(x, y) \to (-x, -y)$.
Therefore, when averaged over one period, feedback from $z$ to $x$ or $y$ (characterized by $Z_x$ or $Z_y$ multiplied by the difference in $z$ components) vanishes and does not help improve the stability of the synchronized state for the coupled Lorenz oscillators. Similarly, feedback from $x$ or $y$ to $z$ (characterized by $Z_z$ multiplied by the difference in $x$ or $y$) does not contribute to the stability.

\section{Summary and discussion}

We have considered a pair of limit-cycle oscillators with weak cross coupling, where different components of the oscillator states are allowed to interact, and optimized the coupling matrix so that the stability of the synchronized state is improved.
For oscillators without frequency mismatch, the optimal coupling matrix yields higher linear stability of the in-phase synchronized state.
For oscillators with frequency mismatch, a range of phase-locked state with given stationary difference can be realized by choosing the coupling matrix appropriately. Necessary conditions for realizability of a given phase difference are also derived.

In this paper, we have derived the optimal coupling matrix that yields the highest linear stability of the synchronized state for linear diffusive coupling given by Eq.~(\ref{model}). This result can be straightforwardly extended to coupled oscillators with general coupling functions, described by
\begin{align}
 \dot{{\bm X}}_1(t) &= {\bm F}_1({\bm X}_1) + \epsilon K {\bm G}({\bm X}_1, {\bm X}_2), \cr
 \dot{{\bm X}}_2(t) &= {\bm F}_2({\bm X}_2) + \epsilon K {\bm G}({\bm X}_2, {\bm X}_1),
\end{align}
where ${\bm G}$ represents general nonlinear coupling between the oscillators $1$ and $2$.
In this case, the phase coupling function in the reduced phase equations~(\ref{phasemodel}) is given by
\begin{align}
\Gamma(\phi) 
&= \frac{1}{2\pi} \int_0^{2\pi} {\bm Z}(\phi + \psi) \cdot  K {\bm G}({\bm X}_0(\phi+\psi), {\bm X}_0(\psi)) d\psi \cr
&= \langle {\bm Z}(\phi + \psi) \cdot K {\bm G}({\bm X}_0(\phi+\psi), {\bm X}_0(\psi)) \rangle_{\psi}
\end{align}
instead of Eq.~(\ref{phscpl0}).
Thus, by defining the function $W(\phi)$ as
\begin{align}
W(\phi) = \langle {\bm Z}(\phi+\psi) \otimes {\bm G} ( {\bm X}_0(\phi+\psi), {\bm X}_0(\psi) ) \rangle_{\psi}
\end{align}
in place of Eq.~(\ref{linearW}) and calculating $V(\phi) = W(\phi) - W(-\phi)$ and $V'(\phi) = W'(\phi) + W'(-\phi)$
from this $W(\phi)$, the optimization can be performed in a similar way to the linear diffusive case. For example, the optimal coupling matrix for the case without frequency mismatch
is given by Eq.~(\ref{KoptcaseA}) with the above $W(\phi)$.

Also, though we have considered only the simple case where all components of the oscillator states can interact with all other components in this paper, it is straightforward to restrict the pairs of components that can actually interact by constraining certain components of $K$ to zero, in order to incorporate realistic physical situations. It would also be interesting to generalize the theory to incorporate different constraints on $K$, for example, 
to reduce the number of non-zero components by assuming sparsity constraint on $K$.

\begin{figure}[htbp]
\centering
\includegraphics[width=0.8\hsize,clip]{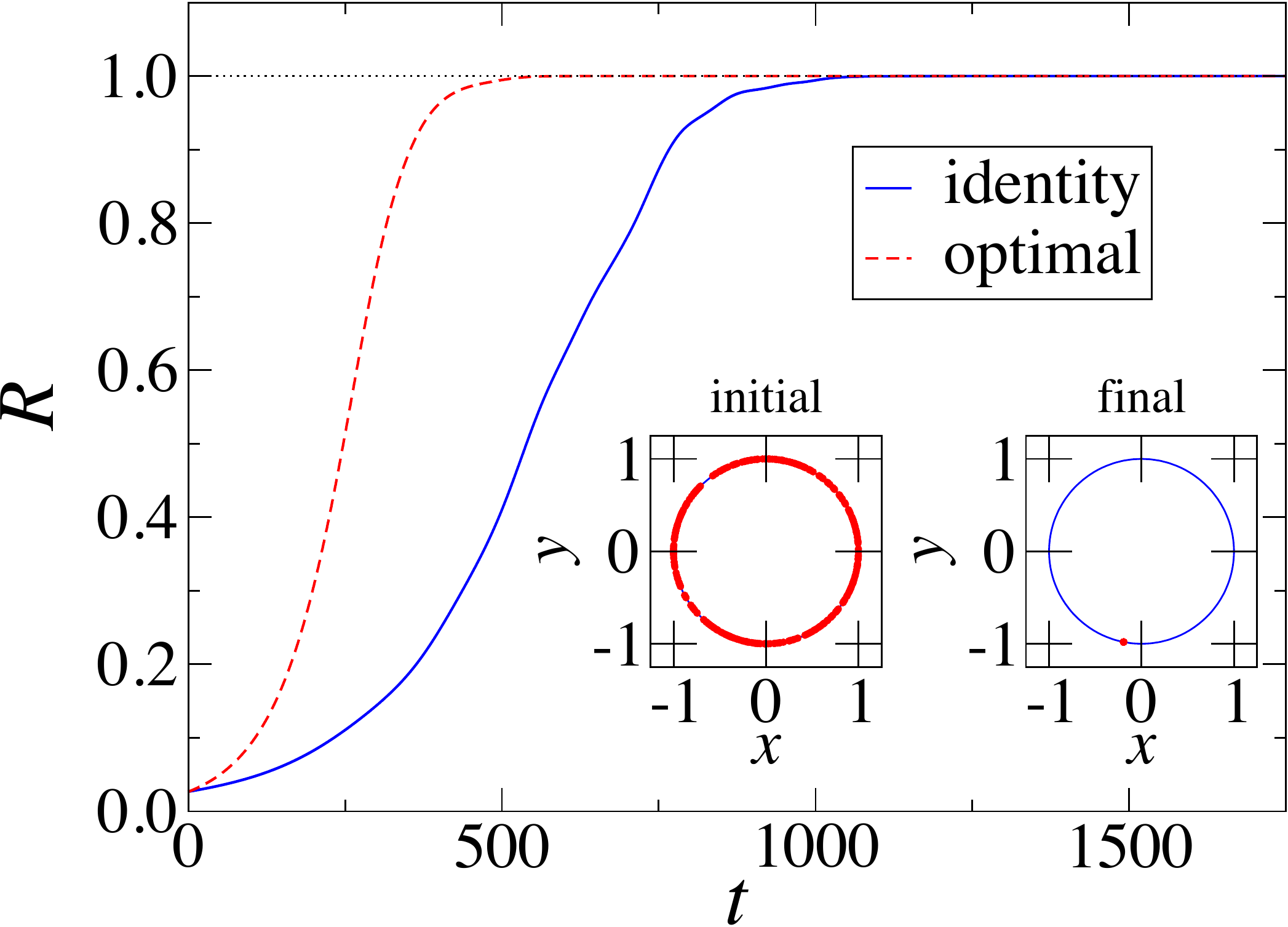}
\caption{Collective synchronization of $N=400$ globally coupled Stuart-Landau oscillators with identical properties. Time sequences of the modulus of the Kuramoto order parameter $R$ for $K=K_I$ and $K=K_{\rm opt}$ are plotted ($P=0.1$ and $\epsilon=0.05$).
The insets show typical snapshots of the oscillator distributions at the initial state and at the final state sufficiently after the convergence.}
\label{fig14}
\end{figure}

Although we have considered only the most fundamental two-oscillator problem in this paper, synchronization of a network of many oscillators have attracted much attention~\cite{Kiss,Kiss2,ref:tanaka08,ref:yanagita10,ref:yanagita12,ref:yanagita14,ref:skardal14,ref:skardal16,ref:nishikawa06a,ref:nishikawa06b,ref:nishikawa10,Ravoori,Dorfler,LiWong,Stankovski}, and generalization of the present framework to many-oscillator networks would be an interesting future problem. For the simplest globally coupled population of $N$ identical oscillators described by
\begin{align}
\dot{\bm X}_i(t) = {\bm F}({\bm X}_i) + \frac{1}{N} \sum_{j=1}^N K ( {\bm X}_j - {\bm X}_i ), \quad i=1, 2, ..., N,
\end{align}
it is expected that the optimal coupling matrix for the two-oscillator case would also provide faster convergence to global synchrony than the identity coupling matrix.
To illustrate this, we simulated $N=400$ SL oscillators with the same parameter values as in Sec. IV, starting from uniformly random initial conditions on the  limit cycle.
Figure~\ref{fig14} shows synchronization processes for $K = K_{\rm opt}$ and for $K = K_I$ in Eqs.~(\ref{glkopt}) and (\ref{identity}), where evolution of the modulus of the Kuramoto order parameter, estimated by $R = \left| (1/N) \sum_{i=1}^N \exp[ \sqrt{-1} \arctan( y_i / x_i ) ] \right|$, is plotted.
We can observe that
the oscillators exhibit much faster convergence to complete synchrony ($R=1$) with $K = K_{\rm opt}$ than with $K = K_I$, as expected.
Of course, for more complex oscillator networks with frequency heterogeneity and coupling randomness, the result of optimization
for the two-oscillator case would not apply due to many-body effects and further investigation should be necessary.

Finally, synchronization between spatiotemporal rhythms in chemical systems has been studied recently~\cite{ref:mikhailov06,ref:mikhailov13,ref:mikhailov14,hildebrand,Fukushima,Epstein}, and generalization of the phase reduction theory to reaction-diffusion equations exhibiting stable spatiotemporal oscillations has also been performed~\cite{Nakao}.
The present framework can also be extended to such situations and can be used to derive the optimal coupling schemes between two coupled spatiotemporal oscillations. A study in this direction is reported in our forthcoming article~\cite{kawamura}, where improvement in the stability of synchronized states between reaction-diffusion systems by introducing linear spatial filters into mutual coupling is considered.

\begin{acknowledgments}

S.S. acknowledges financial support from JSPS (Japan) KAKENHI Grant Number JP15J12045.
Y.K. acknowledges financial support from JSPS (Japan) KAKENHI Grant Number JP16K17769.
H.N. acknowledges financial support from JSPS (Japan) KAKENHI Grant Numbers JP16H01538, JP16K13847, and JP17H03279.

\end{acknowledgments}

\section*{Appendix}

\subsection{Matrix formulas}

The tensor product of $m$-dimensional vectors ${\bm a} = (a_1, ..., a_m)^{\rm T}$ and ${\bm b} = (b_1, ..., b_m)^{\rm T}$ gives a $m \times m$ matrix whose $(i, j)$-component is $[ {\bm a} \otimes {\bm b} ]_{ij} = a_i b_j$ for $i, j = 1, ..., m$.
The inner product of $m \times m$ matrices $A$ and $B$ is defined as
\begin{align}
{\rm Tr\ } (A B^{\rm T}) = \sum_{i=1}^m \sum_{j=1}^m A_{ij} B_{ij} = {\rm Tr\ }( B A^{\rm T} ),
\end{align}
where $A_{ij}$ and $B_{ij}$ represent $(i, j)$-components of the matrices $A$ and $B$, respectively.
The Frobenius norm of a $m \times m$ matrix $A$ is defined as
\begin{align}
\| A \| = \sqrt{ {\rm Tr\ } (A A^{\rm T} ) } = \sqrt{ \sum_{i=1}^m \sum_{j=1}^m A_{ij}^2 },
\end{align}
and the inner product of the matrices $A$ and $B$ is defined as
\begin{align}
{\rm Tr\ } (A B^{\rm T}) = \sum_{i=1}^m \sum_{j=1}^m A_{ij} B_{ij} = {\rm Tr\ }( B A^{\rm T} ).
\end{align}
Derivative of the inner product of matrices is given by
\begin{align}
\frac{d}{dA} {\rm Tr\ } (A B^{\rm T}) = B,
\end{align}
and derivative of the Frobenius norm is given by
\begin{align}
\frac{d}{dA} \| A \|^2 = \frac{d}{dA} {\rm Tr\ } ( A A^{\rm T}) = 2 A.
\end{align}
For arbitrary matrices $A$ and $B$, the Schwartz inequality
\begin{align}
\| A \|^2 \| B \|^2 \geq [ {\rm Tr\ } ( A B^{\rm T}) ]^2
\end{align}
holds, which can be shown by plugging $\lambda = {\rm Tr\ } ( A B^{\rm T}) / \| B \|^2$ into
an inequality $\| A - \lambda B \|^2 \geq 0$ that holds for arbitrary $\lambda$.

\subsection{Calculation of $W(\phi)$ and $V(\phi)$}

Using ${\bm Z}(\theta)$ and ${\bm X}_0(\theta)$, $W(\phi)$ is explicitly given as
\begin{align}
W(\phi) = \langle {\bm Z}(\phi+\psi) \otimes \{ {\bm X}_0(\psi) - {\bm X}_0(\phi+\psi) \} \rangle_{\psi}.
\end{align}
From this $W(\phi)$, the function $V(\phi)$ can be calculated as
\begin{align}
V(\phi) =& W(\phi) - W(-\phi) \cr
=& \langle {\bm Z}(\phi+\psi) \otimes \{ {\bm X}_0(\psi) - {\bm X}_0(\phi+\psi) \} \rangle_{\psi}
\cr
&-
\langle {\bm Z}(-\phi+\psi) \otimes \{ {\bm X}_0(\psi) - {\bm X}_0(-\phi+\psi) \} \rangle_{\psi}
\cr
=& \langle ( {\bm Z}(\phi+\psi) - {\bm Z}(-\phi+\psi) ) \otimes {\bm X}_0(\psi) \} \rangle_{\psi}
\cr
&
-\langle {\bm Z}(\phi+\psi) \otimes {\bm X}_0(\phi+\psi) \rangle_{\psi}
\cr
&+\langle {\bm Z}(-\phi+\psi) \otimes {\bm X}_0(-\phi+\psi) \rangle_{\psi}
\cr
=& \langle ( {\bm Z}(\phi+\psi) - {\bm Z}(-\phi+\psi) ) \otimes {\bm X}_0(\psi) \} \rangle_{\psi},
\end{align}
where $2\pi$-periodicity of the functions ${\bm Z}(\theta)$ and ${\bm X}_0(\theta)$ was used.
Similarly, the derivatives of $W(\phi)$ and $W(-\phi)$ can be calculated as
\begin{align}
W'(\phi) &= \left. \frac{d}{d\psi} W(\psi)\right|_{\psi=\phi} \cr
=& \left \langle {\bm Z}'(\phi+\psi) \otimes \{ {\bm X}_0(\psi) - {\bm X}_0(\phi+\psi) \} \right \rangle_{\psi}
\cr
&- \left \langle {\bm Z}(\phi+\psi) \otimes {\bm X}_0'(\phi+\psi) \right \rangle_{\psi} \cr
=& \langle {\bm Z}'(\phi+\psi) \otimes {\bm X}_0(\psi) \rangle_{\psi}
\cr
&- \left \langle \left( {\bm Z}(\phi+\psi) \otimes {\bm X}_0(\phi+\psi) \right)'\right \rangle_{\psi} \cr
=& \langle {\bm Z}'(\phi+\psi) \otimes {\bm X}_0(\psi) \rangle_{\psi},
\end{align}
and
\begin{align}
W'(-\phi) = \left. \frac{d}{d\psi} W(\psi)\right|_{\psi=-\phi} = \langle {\bm Z}'(-\phi+\psi) \otimes {\bm X}_0(\psi) \rangle_{\psi},
\end{align}
where $2\pi$-periodicity was used again.
Therefore, the derivative of $V(\phi)$ can be calculated as
\begin{align}
V'(\phi) &= W'(\phi) + W'(-\phi) 
\cr
&= \langle ( {\bm Z}'(\phi+\psi) + {\bm Z}'(-\phi+\psi) ) \otimes {\bm X}_0(\psi) \rangle_{\psi}.
\end{align}


\begin{thebibliography}{99}

\bibitem{ref:pikovsky01}
  A.~Pikovsky, M.~Rosenblum, and J.~Kurths,
  {\it Synchronization: A Universal Concept in Nonlinear Sciences}
  (Cambridge University Press, Cambridge, 2001).
  
\bibitem{ref:strogatz03}
  S.~H.~Strogatz,
  {\it Sync: How Order Emerges from Chaos in the Universe, Nature, and Daily Life}
  (Hyperion Books, New York, 2003).

\bibitem{strogatz15}
  S.~H.~Strogatz,
  {\it Nonlinear Dynamics and Chaos, 2nd ed.}
  (Westview Press, Boulder, 2015).

\bibitem{ref:winfree80}
  A.~T.~Winfree,
  {\it The Geometry of Biological Time}
  (Springer, New York, 1980; Springer, Second Edition, New York, 2001).

\bibitem{Kuramoto}
  Y.~Kuramoto,
  {\it Chemical Oscillations, Waves, and Turbulence}
  (Springer, New York, 1984; Dover, New York, 2003).
  
\bibitem{ref:hoppensteadt97}
  F.~C.~Hoppensteadt and E.~M.~Izhikevich,
  {\it Weakly Connected Neural Networks}
  (Springer, New York, 1997).
  
\bibitem{ref:ermentrout10}
  G.~B.~Ermentrout and D.~H.~Terman,
  {\it Mathematical Foundations of Neuroscience}
  (Springer, New York, 2010).


\bibitem{Kiss}
I. Z. Kiss, Y. Zhai, and J. L. Hudson,
Emerging coherence in a population of chemical oscillators,
Science {\bf 296}, 1676 (2002).

\bibitem{Kiss2}
M. Wickramasinghe and I. Z. Kiss,
Spatially organized dynamical states in chemical oscillator networks: Synchronization, dynamical differentiation, and chimera patterns,
PLOS ONE {\bf 8}, e80586 (2013).

\bibitem{ref:tanaka08}
  T.~Tanaka and T.~Aoyagi,
  Optimal weighted networks of phase oscillators for synchronization,
  Phys. Rev. E {\bf 78}, 046210 (2008).
  
\bibitem{ref:yanagita10}
  T.~Yanagita and A.~S.~Mikhailov,
  Design of easily synchronizable oscillator networks using the Monte Carlo optimization method,
  Phys. Rev. E {\bf 81}, 056204 (2010).
  
\bibitem{ref:yanagita12}
  T.~Yanagita and A.~S.~Mikhailov,
  Design of oscillator networks with enhanced synchronization tolerance against noise,
  Phys. Rev. E {\bf 85}, 056206 (2012).
  
\bibitem{ref:yanagita14}
  T.~Yanagita and T.~Ichinomiya,
  Thermodynamic characterization of synchronization-optimized oscillator networks,
  Phys. Rev. E {\bf 90}, 062914 (2014).
  
\bibitem{ref:skardal14}
  P.~S.~Skardal, D.~Taylor, and J.~Sun,
  Optimal synchronization of complex networks,
  Phys. Rev. Lett. {\bf 113}, 144101 (2014).
  
\bibitem{ref:skardal16}
  P.~S.~Skardal, D.~Taylor, and J.~Sun,
  Optimal synchronization of directed complex networks,
  Chaos {\bf 26}, 094807 (2016).

\bibitem{ref:nishikawa06a}
  T.~Nishikawa and A.~E.~Motter,
  Synchronization is optimal in non-diagonalizable networks,
  Phys. Rev. E {\bf 73}, 065106(R) (2006).
  
\bibitem{ref:nishikawa06b}
  T.~Nishikawa and A.~E.~Motter,
  Maximum performance at minimum cost in network synchronization,
  Physica D {\bf 224}, 77 (2006).
  
\bibitem{ref:nishikawa10}
  T.~Nishikawa and A.~E.~Motter,
  Network synchronization landscape reveals
  compensatory structures, quantization, and the positive effect of negative interactions,
  Proc. Natl. Acad. Sci. USA {\bf 107}, 10342 (2010).

\bibitem{Ravoori}
B. Ravoori, A. B. Cohen, J. Sun, A. E. Motter, T. E. Murphy, and R. Roy,
Robustness of optimal synchronization in real networks,
Phys. Rev. Lett. {\bf 107}, 034102 (2011).

\bibitem{Dorfler}
F. D\"orfler and F. Bullo,
Synchronization in complex networks of phase oscillators: a survey,
Automatica {\bf 50}, 1539–1564 (2014).

\bibitem{LiWong}
B. Li
, and K. Y. M. Wong,
Optimizing synchronization stability of the Kuramoto model in complex networks and power grids,
Phys. Rev. E {\bf 95}, 012207 (2017).

\bibitem{Stankovski}
T. Stankovski, T. Pereira, P. V. E. McClintock, and A. Stefanovska,
Coupling functions: universal insights into dynamical interaction mechanisms.
https://arxiv.org/abs/1706.01810,
to appear in Reviews of Modern Physics.

\bibitem{ref:brown04}
  E.~Brown, J.~Moehlis, and P.~Holmes,
  On the phase reduction and response dynamics of neural oscillator populations,
  Neural Comput. {\bf 16}, 673 (2004).

\bibitem{ref:nakao16}
  H.~Nakao,
  Phase reduction approach to synchronization of nonlinear oscillators,
  Contemp. Phys. {\bf 57}, 188 (2016).
  
\bibitem{ref:ashwin16}
  P.~Ashwin, S.~Coombes, and R.~Nicks,
  Mathematical frameworks for oscillatory network dynamics in neuroscience,
  J. Math. Neurosci. {\bf 6}, 1 (2016).

\bibitem{Kawamura1}
  Y.~Kawamura, H.~Nakao, K.~Arai, H.~Kori, and Y.~Kuramoto,
  Collective phase sensitivity,
  Phys. Rev. Lett. {\bf 101}, 024101 (2008).

\bibitem{Kotani}
	K.~Kotani, I.~Yamaguchi, Y.~Ogawa, Y.~Jimbo, H.~Nakao, and G.~B.~Ermentrout,
	Adjoint method provides phase response functions for delay-induced oscillations,
	Phys. Rev. Lett. {\bf 109}, 044101 (2012).

\bibitem{Novicenko1}
	V.~Novicenko and K.~Pyragas,
	Phase reduction of weakly perturbed limit cycle oscillations in time-delay systems,
	Physica D {\bf 241}, 1090 (2012).

\bibitem{Novicenko2}
	V.~Novicenko and K.~Pyragas,
	Phase-reduction-theory-based treatment of extended delayed feedback control algorithm in the presence of a small time delay mismatch,
	Phys. Rev. E {\bf 86}, 026204 (2012).

\bibitem{Nakao}
  H.~Nakao, T.~Yanagita, and Y.~Kawamura,
  Phase-reduction approach to synchronization of spatiotemporal rhythms in reaction-diffusion systems,
  Phys. Rev. X {\bf 4}, 021032 (2014).
  
\bibitem{Kawamura2} 
  Y.~Kawamura and H.~Nakao,
  Collective phase description of oscillatory convection,
  Chaos {\bf 23}, 043129 (2013).
  Y.~Kawamura and H.~Nakao,
  Phase description of oscillatory convection with a spatially translational mode,
  Physica D {\bf 295-296}, 11 (2015).
  
\bibitem{Shirasaka}
	S.~Shirasaka, W.~Kurebayashi, and H.~Nakao,
	Phase reduction theory for hybrid nonlinear oscillators
	Phys. Rev. E {\bf 95}, 012212 (2017).	  

\bibitem{ref:moehlis06}
  J.~Moehlis, E.~Shea-Brown, and H.~Rabitz,
  Optimal inputs for phase models of spiking neurons,
  J. Comput. Nonlin. Dyn. {\bf 1}, 358 (2006).
  
\bibitem{ref:harada10}
  T.~Harada, H.-A.~Tanaka, M.~J.~Hankins, and I.~Z.~Kiss,
  Optimal waveform for the entrainment of a weakly forced oscillator,
  Phys. Rev. Lett. {\bf 105}, 088301 (2010).
  
\bibitem{ref:dasanayake11}
  I.~Dasanayake and J.-S.~Li,
  Optimal design of minimum-power stimuli for phase models of neuron oscillators,
  Phys. Rev. E {\bf 83}, 061916 (2011).
  
\bibitem{ref:zlotnik12}
  A.~Zlotnik and J.-S.~Li,
  Optimal entrainment of neural oscillator ensembles,
  J. Neural Eng. {\bf 9}, 046015 (2012).
  
\bibitem{ref:zlotnik13}
  A.~Zlotnik, Y.~Chen, I.~Z.~Kiss, H.~Tanaka, and J.-S.~Li,
  Optimal waveform for fast entrainment of weakly forced nonlinear oscillators,
  Phys. Rev. Lett. {\bf 111}, 024102 (2013).
  
\bibitem{ref:tanaka14a}
  H.-A.~Tanaka,
  Synchronization limit of weakly forced nonlinear oscillators,
  J. Phys. A: Math. Theor. {\bf 47}, 402002 (2014).
  
\bibitem{ref:tanaka14b}
  H.-A.~Tanaka,
  Optimal entrainment with smooth, pulse, and square signals in weakly forced nonlinear oscillators,
  Physica D {\bf 288}, 1 (2014).
  
\bibitem{ref:hasegawa14a}
  Y. Hasegawa and M. Arita,
  Circadian clocks optimally adapt to sunlight for reliable synchronization,
  Journal of the Royal Society Interface 11, 20131018 (2014).

\bibitem{ref:hasegawa14b}
  Y. Hasegawa and M. Arita,
  Optimal implementations for reliable circadian clocks,
  Phys. Rev. Lett. 113, 108101 (2014).

\bibitem{ref:pikovsky15}
   A.~Pikovsky,
   Maximizing coherence of oscillations by external locking,
   Phys. Rev. Lett. {\bf 115}, 070602 (2015).
         
\bibitem{ref:tanaka15}
  H.-A.~Tanaka, I.~Nishikawa, J.~Kurths, Y.~Chen, and I.~Z.~Kiss,
  Optimal synchronization of oscillatory chemical reactions
  with complex pulse, square, and smooth waveforms signals maximizes Tsallis entropy,
  Europhys. Lett. {\bf 111}, 50007 (2015).
  
\bibitem{ref:zlotnik2016}
  A.~Zlotnik, R.~Nagao, I.~Z.~Kiss, and J.-S.~Li,
  Phase-selective entrainment of nonlinear oscillator ensembles,
  Nature Communications {\bf 7}, 10788 (2016).

\bibitem{ref:mikhailov06}
  A.~S.~Mikhailov and K.~Showalter,
  Control of waves, patterns and turbulence in chemical systems,
  Phys. Rep. {\bf 425}, 79 (2006).
  
\bibitem{ref:mikhailov13}
  A.~S.~Mikhailov and G.~Ertl (Editors),
  {\it Engineering of Chemical Complexity}
  (World Scientific, Singapore, 2013).
  
\bibitem{ref:mikhailov14}
  A.~S.~Mikhailov and G.~Ertl (Editors),
  {\it Engineering of Chemical Complexity II}
  (World Scientific, Singapore, 2014).

\bibitem{Epstein}
	I.~R.~Epstein, I.~B.~Berenstein, M.~Dolnik, V.~K.~Vanag, L.~Yang, A.~M.~ 	Zhabotinsky,
	Coupled and forced patterns in reaction-diffusion systems,
	Phil. Trans. R. Soc. A {\bf 366}, 397–408 (2008).
	
\bibitem{hildebrand}
	M.~Hildebrand, J.~Cui, E.~Mihaliuk, J.~Wang, and K.~Showalter,
	Synchronization of spatiotemporal patterns in locally coupled excitable media,
	Phys. Rev. E {\bf 68}, 026205 (2003).
	
\bibitem{Fukushima}
	S. Fukushima, S. Nakanishi, K. Fukami, S. I. Sakai, T. Nagai, T. Tada, and Y. Nakato,
	Observation of synchronized spatiotemporal reaction waves in coupled 	electrochemical oscillations of an NDR type,
	Electrochem. Comm. {\bf 7}, 411 (2005). 

\bibitem{kawamura}
	Y.~Kawamura, S.~Shirasaka, T.~Yanagita, and H.~Nakao,
	Optimizing mutual synchronization of rhythmic spatiotemporal patterns in reaction-diffusion systems,
	Phys. Rev. E (2017) (to be published).

\bibitem{footnote}
Although not the focus of the present paper, for networks of many elements, optimization of coupling topology has also been studied extensively for phase oscillators~\cite{ref:tanaka08,ref:yanagita10,ref:yanagita12,ref:yanagita14,ref:skardal14,ref:skardal16,Dorfler,LiWong}, where the Kuramoto order parameter is typically used as a quantifier of synchrony, and for chaotic oscillators~\cite{ref:nishikawa06a,ref:nishikawa06b,ref:nishikawa10,Ravoori}, where maximization of the range of coupling intensity under the framework of the master stability function is often considered.

\bibitem{footnote2}
In this particular example of coupled Brusselators, $\Gamma_a(\phi)$ for $K_I$ is close to a sinusoidal curve as shown in Fig.~\ref{fig05}, and it yields a small stationary phase difference $\phi^* = 0.378$ when $\Delta \omega = 0.175$  (note that $\Gamma_a(\phi)$ for $K_I$ is the same irrespective of $\Delta \omega$).
On the other hand, as shown in Fig.~\ref{fig10}, when $\Delta \omega = 0.175$, the closer $\phi^*$ is to $0$, the optimized stability of $\phi^*$ is higher and the functional form of $\Gamma_a(\phi)$ for $K_{\rm opt}$ is (roughly) closer to a sinusoidal curve.
Because the value $\phi^*=0.378$ is fairly close to $0$, $K_I$ yields reasonably high stability close to the value for $K_{\rm opt}$ in this example.
In general, $K_I$ does not necessarily yield such high stability comparable to $K_{\rm opt}$.
  
\end{thebibliography}
\end{document}